\documentclass{emulateapj}
\usepackage{natbib}
\usepackage{multirow}
\usepackage{graphicx}
\shorttitle{Molecular Gas in Circinus}
\shortauthors{Zschaechner, et al.}

\begin{document}

\title{The Molecular Wind in the Nearest Seyfert Galaxy Circinus Revealed by ALMA} 
\author{\sc Laura K. Zschaechner\altaffilmark{1}, Fabian Walter\altaffilmark{1}, Alberto Bolatto\altaffilmark{2}, Emanuele P. Farina\altaffilmark{1}, J. M. Diederik Kruijssen\altaffilmark{3, 1}, Adam Leroy\altaffilmark{4}, David S. Meier\altaffilmark{5},J\"{u}rgen Ott\altaffilmark{6}, and Sylvain Veilleux\altaffilmark{2}}
\altaffiltext{1}{Max Planck Institute f\"{u}r Astronomie - K\"{o}nigstuhl 17, 69117 Heidelberg - Germany; zschaechner@mpia.de}  
\altaffiltext{2}{Department of Astronomy and Joint Space Science Institute, University of Maryland, College Park, MD 20642, USA}
\altaffiltext{3}{Astronomisches Rechen-Institut, Zentrum f\"{u}r Astronomie der Universit\"{a}t Heidelberg, M\"{o}nchhofstra{\ss}e 12-14, D-69120 Heidelberg, Germany}
\altaffiltext{4}{Department of Astronomy, The Ohio State University, 140 West 18th Avenue, Columbus, OH 43210, USA; National Radio Astronomy Observatory, 520 Edgemont Road, Charlottesville, VA 22903, USA}
\altaffiltext{5}{Department of Physics, New Mexico Institute of Mining and Technology, 801 Leroy Place, Socorro, NM 87801, USA}
\altaffiltext{6}{National Radio Astronomy Observatory - P.O. Box O, 1003 Lopezville Road, Socorro, NM 87801, USA}

\slugcomment{Accepted to The Astrophysical Journal on September 14, 2016}

\begin{abstract}

\par
   We present ALMA observations of the inner 1' (1.2 kpc) of the Circinus galaxy, the nearest Seyfert. We target CO (1--0) in the region associated with a well-known multiphase outflow driven by the central active galactic nucleus (AGN).  While the geometry of Circinus and its outflow make disentangling the latter difficult, we see indications of outflowing molecular gas at velocities consistent with the ionized outflow.  We constrain the mass of the outflowing molecular gas to be 1.5$\times$10$^{5}$ --5.1$\times$10$^{6}$ M$_{\odot}$, yielding a molecular outflow rate of 0.35--12.3 M$_{\odot}$ yr$^{-1}$.  The values within this range are comparable to the star formation rate in Circinus, indicating that the outflow indeed regulates star formation to some degree.  The molecular outflow in Circinus is considerably lower in mass and energetics than previously-studied AGN-driven outflows, especially given its high ratio of AGN luminosity to bolometric luminosity.  The molecular outflow in Circinus is, however, consistent with some trends put forth in \citet{2014A&A...562A..21C}, including a linear relation between kinetic power and AGN luminosity, as well as its momentum rate vs.\ bolometric luminosity (although the latter places Circinus among the starburst galaxies in that sample).  We detect additional molecular species including CN and C$^{17}$O.

\end{abstract}


\section{Introduction}\label{introduction}

\par
A primary source of feedback in galaxies are galactic outflows/winds \citep{2005ARA&A..43..769V}.  They potentially expel material into the intergalactic medium (IGM), enriching the surrounding environment, and constituting a reservoir of material that can be reaccreted to fuel later star formation in galaxies \citep{2010MNRAS.406.2325O}. Galactic winds play also an important role in driving bubbles into the ISM that allow radiation to propagate further \citep{2000ApJ...531..846D}.  Galactic winds are invoked to explain the observed paucity of massive galaxies in the local universe (e.g.\ \citealt{2008MNRAS.391..481S}) and the quenching of star formation in massive galaxies (e.g.\ \citealt{2013MNRAS.430.3213B}).  Thus, it is imperative to gauge the total outflow rate and compare it to the accretion rate in order to understand the evolution of galaxies.  While a recent emphasis has been placed on researching galactic winds (e.g.\ \citealt{2005ARA&A..43..769V}, \citealt{2010ApJ...711..818S}, \citealt{2010A&A...518L.155F}, \citealt{2013Natur.499..450B}, \citealt{2014A&A...562A..21C}, \citealt{2015ApJ...798...31A}, \citealt{2015A&A...574A..85A}), the degree to which they regulate these evolutionary processes is still poorly constrained, largely due to insufficient observational data.  With the advent of instruments such as ALMA, the molecular phases of galactic winds may now be mapped in detail and with superb velocity resolution.

\par
A fundamental question is the relative importance of starburst- and AGN-driven winds in the feedback process.   Both tend to suppress star formation by removing molecular gas, but the efficiency with which they do it is unknown.  Both affect the surrounding ISM, but their respective impacts may differ based on their energetics and distributions within the host galaxies.  Simulations and observations indicate that AGN-driven winds are more energetic and concentrated near the center, resulting in a higher chance for material to escape the galaxy and a lower likelihood of it fueling future star formation \citep{2012ApJ...750...55S}. Starburst-driven winds, however, are more widely distributed and may eject material at larger galactocentric radii, producing a greater effect on the ISM. When both are present, the two types of outflows may work to amplify their respective contributions \citep{2009MNRAS.396L..46P}, but can also counteract each other's effects \citep{2013NatSR....E1738B}. Fundamentally, their impact depends on their efficiency at entraining and ejecting material of the various phases of the ISM from in the disk.

\par
    Molecular outflows are seen in several starburst galaxies, including M 82 (\citealt{2002ApJ...580L..21W}, \citealt{2015ApJ...814...83L}), Arp 220 \citep{2009ApJ...700L.104S}, NGC 253 \citep{2013Natur.499..450B}, NGC 1808 \citep{2016ApJ...823...68S} among others.  The molecular mass outflow rates in these galaxies indicate that starburst-driven molecular outflows can indeed regulate star formation (SF), particularly in NGC 253 where the molecular mass outflow rate was determined to be $>$3 times the global star formation rate (SFR) in that galaxy, and likely higher.

\par
   During the past five years there have been a number of studies focusing on AGN-driven outflows, summarized in \citet{2011ApJ...733L..16S}, \citet{2013ApJ...776...27V}, \citet{2014A&A...562A..21C}, \citet{2015A&A...583A..99F}, and \citet{2015ApJ...811...15L}.  These studies show that an increase in AGN luminosity (L$_{AGN}$) typically results in stronger outflows, with some reaching values of over 1000 M$_{\odot}$ yr$^{-1}$ for the molecular component.  There are also indications that stronger outflows may also be tied to a high ratio of L$_{AGN}$ to the bolometric luminosity (L$_{bol}$), but that appears to be less important. 

\par

    Other results by \citet{2015ApJ...798...31A} indicate a molecular outflow of $\sim$100 M$_{\odot}$ yr$^{-1}$ with an average velocity of 177 km s$^{-1}$ within a radius of $\sim$225 pc in NGC 1266, and \citet{2016A&A...590A..73A} determine a molecular mass outflow rate for NGC 1377 of 9--30 M$_{\odot}$ yr$^{-1}$, showing a strong jet potentially entraining the cold gas extending out to $\sim$150 pc in projection, with velocity estimates from 240 to 850 km s$^{-1}$.  \citet{2015A&A...580A...1M} and \citet{2014Natur.511..440T} characterize the AGN-driven molecular outflow in IC 5063, a radio-weak Seyfert galaxy, that the multiphase outflow components (ionized, {\sc H\,i}, warm H$_{2}$, cold molecular gas) share similar kinematics.  The molecular mass outflow rate they determine is sufficiently high and the AGN sufficiently weak, that the radio jet within that galaxy is likely the primary driver of the outflow - not radiation from the AGN.  However, in Mrk 231, \citet{2016A&A...593A..30M} find that it is unlikely that a jet is driving its multiphase ({\sc Na\,i}, {\sc H\,i}, OH, and CO) outflow.


\par
In contrast with those massive molecular outflows, \citet{2007A&A...468L..49M} detect a more modest molecular outflow of $\sim$4 M$_{\odot}$ yr$^{-1}$ in M 51 - a galaxy that is neither starbursting nor a ULIRG.  They conclude that the molecular outflow is likely mechanically-driven by a radio jet as opposed to SF (which is spread throughout the disk) or radiation from the AGN (which is too weak).  The bulk of the molecular outflow they detect is within 1" (34 pc) of the center of M 51 -- quite small by comparison to the molecular outflows observed in other systems and unlikely to impact SF throughout M 51.  \citet{2014A&A...565A..97C} observe NGC 1566, a galaxy with a low-luminosty AGN, with ALMA and find no evidence for a molecular outflow or AGN-driven feedback.

\par
  In the current sample of molecular outflows there appears to be a correlation between AGN luminosity and powerful outflows.  There is however, (as noted in \citealt{2014A&A...562A..21C}), a bias in that, with a few exceptions (e.g.\ ULIRGs in the Herschel sample presented in \citealt{2010A&A...518L..36S} for which the molecular outflows were investigated by \citealt{2013ApJ...776...27V}), only galaxies with previously-detected, high-velocity molecular outflows are considered.  It is necessary to observe molecular outflows in host galaxies with a range of AGN luminosities at high spatial resolution in order to see if these trends hold.

\subsection{The Circinus Galaxy}
\par
\begin{figure}
\includegraphics[width=80mm]{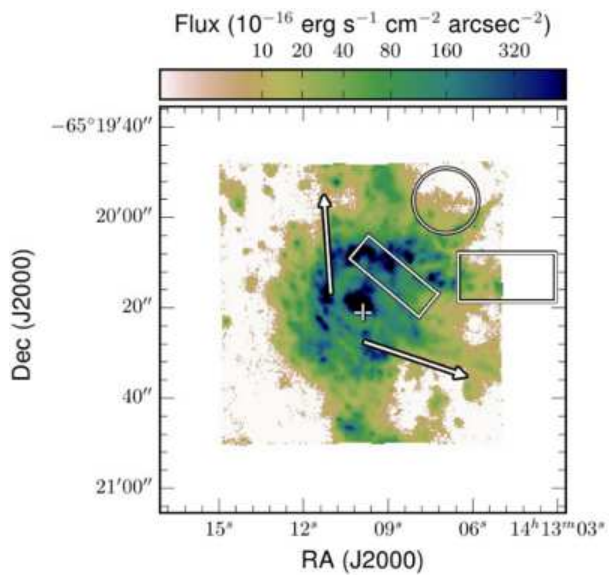}
\caption{\textit{An H$\alpha$ image of Circinus created with MUSE data (full presentation of the MUSE data in Venturi at al. (\textit{in prep}) and Marconi et al. (\textit{in prep})). Multiple filaments extend from the central regions of the disk (center marked with a cross).  The approximate width of the H$\alpha$ outflow cone is shown by the two large arrows.  Regions referred to throughout the text are introduced here for reference although their molecular components are not visible.  The slanted rectangle near the center marks the molecular ``overdense region", the circle marks the ``NW cloud", and the rectangle to the west marks the ``far W cloud".  Assuming our adopted distance of 4.2 Mpc, 1" = 20.4 pc.} \label{halpha_muse}}
\end{figure}

   At an adopted distance of 4.2 Mpc (1"$=$20.4 pc, \citealt{2009AJ....138..323T}), the nearest Seyfert 2 \citep{1994A&A...288..457O} is The Circinus galaxy \citep{1977A&A....55..445F}.   An ionized outflow is clearly observed emerging from the nuclear region of Circinus, with blue-shifted velocities (receding components are likely obscured by the intervening disk) of approximately $-$150 km~s$^{-1}$ (e.g.\ \citealt{1994Msngr..78...20M}, \citealt{1997ApJ...479L.105V}, \citealt{2000AJ....120.1325W}, Figure~\ref{halpha_muse}). The ionized component extends for at least 30" along the minor axis and farther.  \citet{1997ApJ...479L.105V} also observe what look like bow-shocked features resembling Herbig-Haro objects at the ends of these filaments suggesting a strong interaction with the surrounding ISM. \citet{1998MNRAS.297.1202E} detect an outflow in the radio continuum, most prominently in the north-eastern direction, but with a counter-jet to the southeast.  This counter-jet shows that the outflow is not entirely one-sided, and observations of the molecular phase (which suffers less from extinction effects) may indeed uncover a counter-jet closer to the disk -- a possibility we explore in this work. Observational parameters of Circinus are listed in Table~\ref{tbl_circ}.  

\begin{deluxetable}{lcc}
\tabletypesize{\scriptsize}
\tablecaption{The Circinus Galaxy  \label{tbl_circ}}
\tablewidth{0pt}
\tablehead
{
\colhead{Parameter} &
\colhead{Value} &
\colhead{Reference}
}
\startdata
\phd Distance &4.2 Mpc&\citet{2009AJ....138..323T}\\
\phd SFR &4.7 M$_{\odot}$ yr$^{-1}$&\citet{2008ApJ...686..155Z}\tablenotemark{a}\\
\phd L$_{bol}$ &1.7$\times$10$^{10}$ L$_{\odot}$ &Maiolino et al.\ (1998)\tablenotemark{b}\\
\phd L$_{AGN}$ &10$^{10}$ L$_{\odot}$ &Moorwood et al.\ (1996)\\
\phd Type &Seyfert 2&Oliva et al.\ (1994)\\
\phd Black Hole Mass &1.7$\times$10$^{6}\pm0.3$M$_{\odot}$&Tristram et al.\ (2007)\\
\phd Eddington Luminosity &5.6$\times$10$^{10}$L$_{\odot}$&Tristram et al.\ (2007)

\tablenotetext{a}{Multiple values from the literature for the star formation rate are listed in \citet{2012MNRAS.425.1934F}.  We adopt the value derived using L$_{TIR}$.}
\tablenotetext{b}{Originally derived by \citet{1997A&A...325..450S} and modified by \citet{1998ApJ...493..650M}.}

\enddata
\end{deluxetable}

\par
This paper is organized as follows: In $\S$~\ref{observations}--~\ref{data} we present the observations, calibration, and the data.  In $\S$~\ref{results} we present our analysis and results followed by a discussion in $\S$~\ref{discussion}.  We provide a summary of our results in $\S$~\ref{summary}.  The Appendix includes details on our kinematic modeling.

\section{Observations \& Data Reduction}\label{observations}

\par
     In order to study the molecular outflow in Circinus, we mapped the inner 1' (1.2 kpc) in CO (1--0) emission line (rest frequency 115.27120 GHz) with ALMA.  These observations include additional frequencies from 99.8--115.7 GHz.   The velocity resolution of our data is 2.5--3 km s$^{-1}$ over this frequency range.  Our observational setup consisted of a 7-pointing mosaic with the 12-meter array, a 3-pointing mosaic for the ALMA Compact Array (ACA), and single-pointing Total Power TP observations. Observing dates, times, and calibrators are listed in Table~\ref{tbl_1}.

\begin{deluxetable}{lr}
\tabletypesize{\scriptsize}
\tablecaption{Observational and Instrumental Parameters  \label{tbl_1}}
\tablewidth{0pt}
\tablehead
{
\colhead{Parameter} &
\colhead{Value}\\
}
\startdata
\phd Observation Dates$-$ACA &2014 Jul 02\\
\phd \hspace{42pt} &2014 Aug 10\\
\phd \hspace{42pt} &2014 Aug 11\\
\phd Observation Dates$-$12-meter Array &2014 Dec 27\\
\phd Observation Dates$-$Total Power Array &2015 Jan 14 \\
\phd \hspace{42pt} &2015 Jan 15\\
\phd \hspace{42pt} &2015 Jan 17\\
\phd \hspace{42pt} &2015 Jan 27\\
\phd \hspace{42pt} &2015 Jan 31\\

\phd Number of Antennas ACA&10\\
\phd Number of Antennas 12-meter&36\\

\phd Pointing Center &14h 13m 09.9s\\
\phd \hspace{42pt} &-65d 20m 21.0s\\ 

\phd Calibrators ACA - Flux&Mars\\
\phd \hspace{65pt} Gain&J1617--5848\\
\phd \hspace{65pt} Bandpass&J1308--6707\\

\phd Calibrators 12-m - Flux&Ceres\\
\phd \hspace{65pt} Gain&J1227--6509\\
\phd \hspace{65pt} Bandpass&J1427--4206\\

\phd Calibrator TP - All&J1424--6807\\

\phd Channel Spacing - Full bandwidth &976.562 kHz\\
\phd Channel Spacing - Individual Line Cubes &3.0 km s$^{-1}$\\
\phd Total Bandwidth & 7.5 GHz\\
\phd Frequencies Covered & 99.8--103.5 GHz\\
\phd \hspace{42pt} & 111.9--115.7 GHz\\
\phd Spatial Resolution/Beam Size &2--3"\\  
\phd \hspace{42pt} & 40--60 pc\\
\phd RMS Noise (SPW 2 99.8 to 101.7 GHz) &1.8 mJy beam$^{-1}$\\
\phd \hspace{42pt} (SPW 3 101.6 to 103.5 GHz) &1.9 mJy beam$^{-1}$\\
\phd \hspace{42pt} (SPW 1 111.9 to 113.9 GHz) &2.4 mJy beam$^{-1}$\\
\phd \hspace{42pt} (SPW 0 113.7 to 115.7 GHz) &3.0 mJy beam$^{-1}$\\
\phd \hspace{42pt} (CO (1--0) line cube) &5.4 mJy beam$^{-1}$\\  
\phd \hspace{42pt} (CN (1--0), J=1/2 line cube) &2.8 mJy beam$^{-1}$\\
\phd \hspace{42pt} (CN (1--0), J=3/2 line cube) &3.3 mJy beam$^{-1}$

\enddata
\end{deluxetable}

\par
   ALMA data are from programme ID: ADS/JAO.ALMA\#2013.1.00247.S, PI: Zschaechner. Calibration of the ALMA data was done using the Common Astronomy Software Application (CASA).  Initial calibration utilized the scripts provided by the ALMA staff for the ACA data and the ALMA pipeline for the 12-meter data.  Minor adjustments were made to the ACA calibration.  Minor phase errors in the final cubes were eliminated via a single iteration of phase-only self-calibration on the 12-meter data. Amplitude self-calibration was attempted but yielded no clear improvement and was not used. 

\par
Two sets of cubes were created:  one set containing the full bandwidth of each spectral window (full spectral resolution), and another set containing each detected line (3 km s$^{-1}$ spectral resolution).  Cubes were created using both natural and Briggs weighting \citep{1995PhDT.......125B} with a robust parameter of 0.5 in order to maximize sensitivity and resolution, respectively.  Masks based on naturally-weighted cubes with thresholds of 3$\sigma$ were used to create clean boxes for the individual line cubes.   The only exception was for the the CO (1--0) cube which, due to substantially brighter emission and larger radial extent, required a 6$\sigma$ mask (based on the naturally-weighted cube) to remove artefacts.  While using that mask, the final Briggs-weighted cube was cleaned down to 4$\sigma$, which was optimal for detecting the most flux while avoiding additional low-level artefacts.    The cubes were then combined with data from the total power array using {\tt FEATHER}. The final cube parameters are given in Table~\ref{tbl_1}.  The beam size is approximately 3$\times$2" for the naturally-weighted cubes, and 2.7$\times$2.1" for the CO (1--0) Briggs-weighted cube which we present for all CO (1--0) maps.  Representative channel maps of the CO (1--0) line are shown in Figure~\ref{channel_maps}.  

\par
   Integral--field spectroscopic data of the Circinus galaxy were collected during March 11th, 2015 with the Multi Unit Spectroscopic
Explorer (MUSE, \citep{2010SPIE.7735E..08B}) on the ESO/VLT (programme ID: 094.B-0321(A), PI: Marconi). The seeing during the 4$\times$500s science exposures was mostly
sub--arcsecond with small variations between 0\farcs85 to 1\farcs05.  The MUSE data reduction pipeline (version 1.0.1) was used to perform
standard steps of the data processing (\citealt{2012SPIE.8451E..0BW}, \citealt{2014ASPC..485..451W}).   These steps included bias, flat field, illumination corrections, wavelength, and flux calibrations (using observations of the standard star EG274 collected in same night of the science images).  The galaxy fills the MUSE field--of--view (1$^\prime$$\times$1$^\prime$).  Before combining the single exposures, a sky model was created and subtracted from 4$\times$100s exposures of mostly empty sky fields taken between the science images.  Finally, single exposures were sampled on the typical 0\farcs2$\times$0\farcs2$\times$1.25$\AA$ grid and then combined.  Wavelength calibration was refined in post-processing using the centroid of bright sky emission lines. This procedure results in an accuracy of better than 1 tenth of pixel.

\section{The Data}\label{data}

\par
\begin{figure*}
\includegraphics[width=170mm]{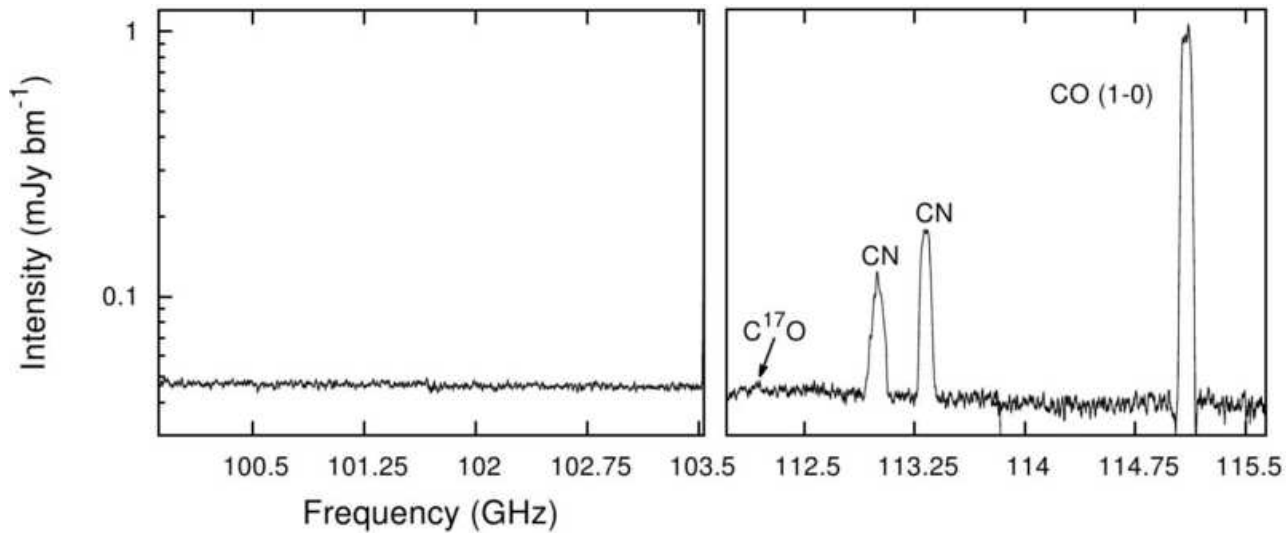}
\caption{\textit{A full spectrum of all four spectral windows taken at the center of Circinus (left: lower sideband, right: upper sideband).  Note the logarithmic scale.  The CO (1--0) line appears prominently near 115 GHz, while the CN (1--0) J=1/2 and J=3/2 hyperfine structure lines are detected near 113 GHz.  C$^{17}$O is also detected at in the center (peaking at 8 mJy bm$^{-1}$ in a 3 km s$^{-1}$ channel).  The average beam for this spectrum is 2.5" and the values shown are taken at the pixel corresponding to the central pointing.} \label{continuum_slope_compact_avg}}  
\end{figure*}
\begin{figure*}
\includegraphics[width=170mm]{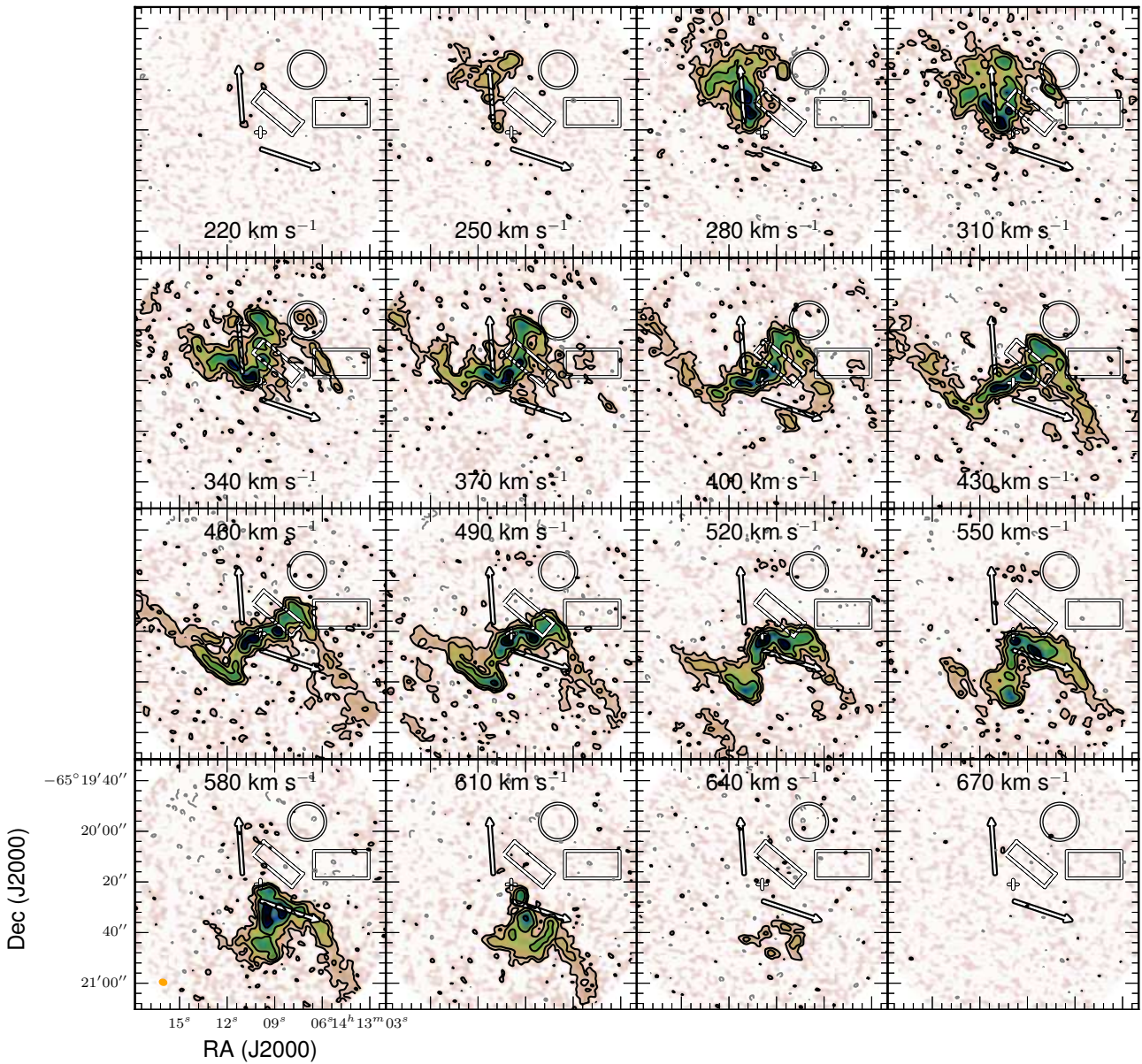}
\caption{\textit{Channel maps of the CO (1--0) at 3 km s$^{-1}$ showing every tenth channel.  Contours begin at 3$\sigma$ and increase by factors of three.  Negative 3$\sigma$ contours are shown in gray.   Marked features are as in Figure~\ref{halpha_muse}.  The beam is shown as the orange ellipse in the lower-left corner.  Note how there exists a large amount of spiral arm material at and near the systemic velocty (434 km s$^{-1}$), indicating substantial streaming.  A potential 3$\sigma$ counter-rotating feature is seen to the north (adjacent to the region containing the NW cloud) from 480--590 km s$^{-1}$.  Deeper observations or observations of additional CO transitions will be able to tell if this feature is real.  The feature traverses towards the center and elongates in the panel corresponding to 590 km s$^{-1}$}.   \label{channel_maps}}
\end{figure*}

    ALMA detects bright molecular emission across Circinus, with a typical spectrum shown in Figure~\ref{continuum_slope_compact_avg}.  The continuum has a slight downward slope with increasing frequency and is on the order of 0.05 mJy bm$^{-1}$ as shown in Figure~\ref{continuum_slope_compact_avg}.  There appear to be no lines detected from 100--103.5 GHz.  Prominent lines appear between 112 GHz and 115.5 GHz, including CO (1--0), CN (1--0) - both J=1/2 and J=3/2 groups of hyperfine transitions, and C$^{17}$O.  A summary of the detected lines is given in Table~\ref{tbl_2}.

\par
\begin{figure*}
\includegraphics[width=170mm]{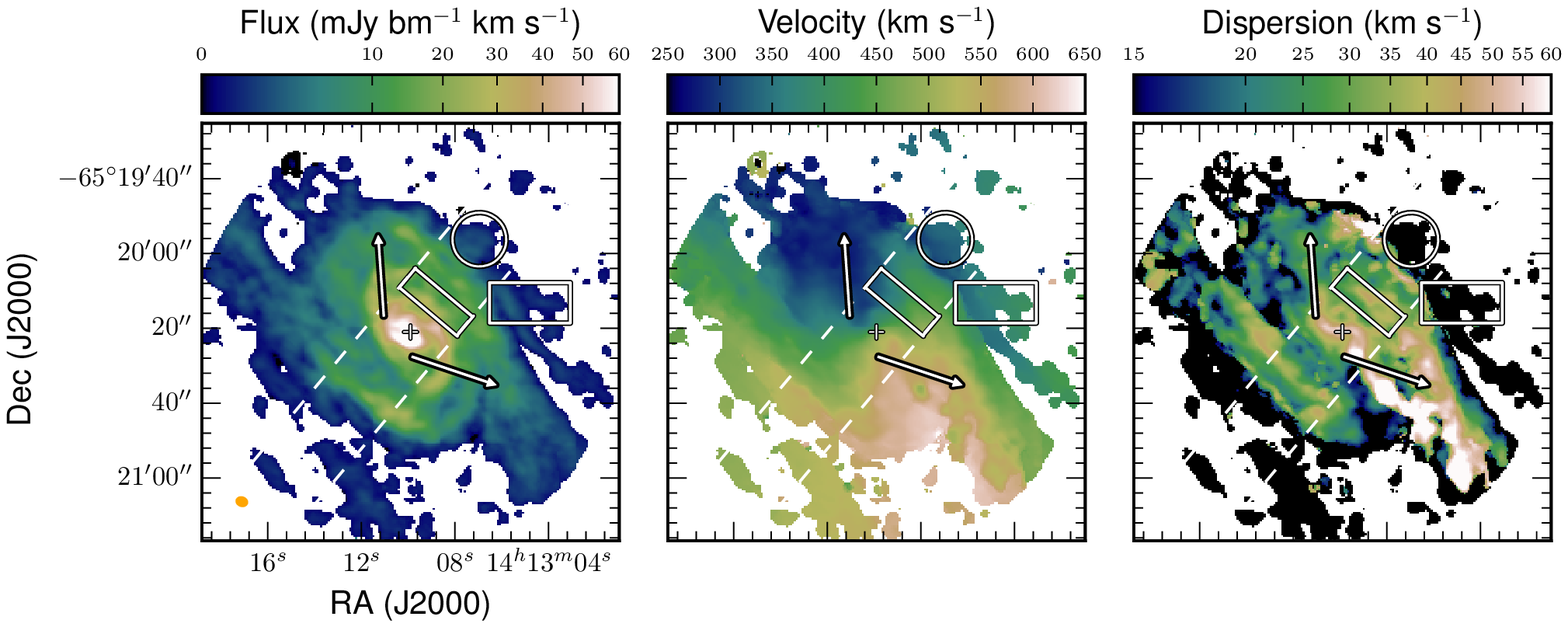}
\caption{\textit{CO (1--0) moment maps of Circinus.  Left:  An integrated intensity (zeroth-moment) map of CO (1--0) in the central $\sim$1' of Circinus. Marked features are as in Figure~\ref{halpha_muse} with the exception of the white dashed lines, which indicate the region considered in Figure~\ref{dispersion_profile}.  Note the overdensity in the northwest region extending $\sim$40" from the nucleus to the northwest.  Center:  The intensity-weighted velocity map (first-moment) of the same region.  Overall, the disk displays normal rotation, with some disturbances to the southwest and northwest.  Right:  An intensity-weighted velocity dispersion (second-moment) map.    Further north there is an abrupt ridge, possibly due to disk and/or outflowing material meeting the northern spiral arm, which has a different inclination from the rest of the disk.   In all cases, the cube is masked at 13.5 mJy bm$^{-1}$ ($\sim$2.5$\sigma$) prior to creating the maps.    The beam is shown in orange in the bottom corner of the left panel.} \label{moments_CO}}
\end{figure*}

   The most prominent line throughout the spectrum is CO (1--0), which is the primary focus of this paper.  Moment maps are created using software described in \citet{2015MNRAS.448.1922S} in order to retain faint emission.  Figure~\ref{moments_CO} (left panel) shows moment maps of the CO (1--0) line.  The integrated intensity (zeroth-moment) map shows a relatively symmetric main disk, with spiral arms extending to the northeast and southwest at both large and small radii.  A moderate overdensity is present in the northwestern half of the galaxy (henceforth referred to as the overdense region).    

\par
   The intensity-weighted velocity (first-moment) map (Figure~\ref{moments_CO}, middle panel) shows that the disk has a well-behaved rotation curve peaking at $\sim$200 km s$^{-1}$ (deprojected), with the primary departures from circular rotation being in the two outer spiral arms and to the south.  Interestingly, to match the observed velocities of the outer spiral arms without introducing radial motions, their rotational velocities would need to go to zero. This effect is likely produced via a combination of projection effects and streaming motions.  

\par
    The intensity-weighted velocity dispersion (second-moment) map (Figure~\ref{moments_CO}, right panel) shows substantial structure throughout the disk.  There is a disturbance to the southwest (henceforth referred to as the ``southern disturbance") with high velocity dispersion that goes approximately along the major axis and to the end of the outer spiral arm.  The two spiral arms show an increased dispersion, especially directly along the edge of the main disk.  To the north there is an abrupt decrease in the velocity dispersion when meeting the spiral arm, as well as a sudden change in velocity (first-moment map).  Naturally, the dispersion near the center is expected to be high, but note here the increased dispersion in the overdense region noted in the integrated intensity map - consistent with a wind driven from the center.  A similar increase in dispersion is not present to the southeast (azimuthally-opposite region).

\subsection{Likely Outflowing Molecular Gas}\label{structure}

\par
\begin{figure*}
\begin{centering}
\includegraphics[width=70mm, angle=270]{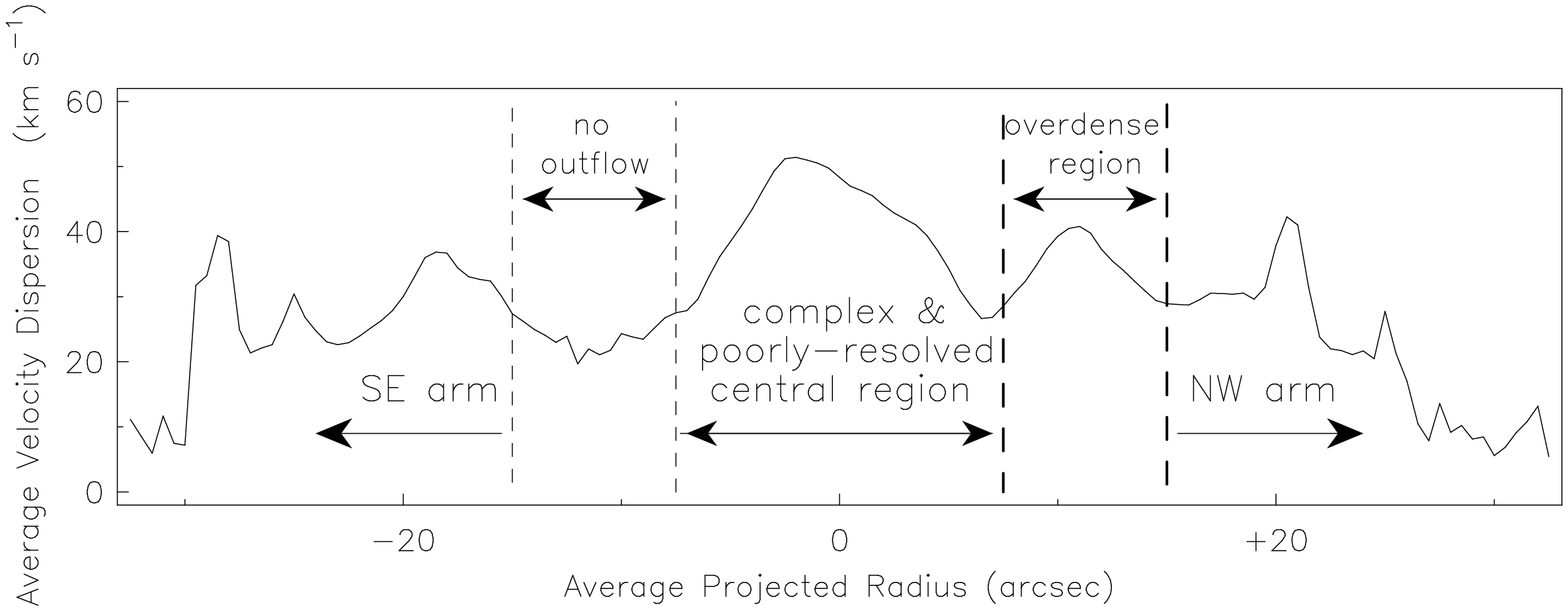}
\caption{\textit{The average velocity dispersion with projected radius for a 20"-wide strip along the region containing the outflow (white dashed lines in Figure~\ref{moments_CO}). The central $\sim$7.5" are poorly-resolved and contain a bar as well as spiral structure.  Thus, they are not included in our estimates of the outflow properties.  The NW and SW arm regions are also excluded.  Considering only the remaining regions, there is a clear factor of approximately two increase in velocity dispersion over the 7.5$\times$20" ``outflow" region containing the overdensity seen in Figure~\ref{moments_CO} as compared to the ``no outflow" region 180$^{\circ}$ to the southeast (which itself has a velocity dispersion typical of a spiral disk).  } \label{dispersion_profile}}
\end{centering}
\end{figure*}

   Most emission that is likely to be part of a molecular outflow is located northwest of the center in the aforementioned overdense region approximately 7.5--15" or 330--650 pc as it extends along the minor axis (Figure~\ref{moments_CO}).   The molecular outflow itself is most readily seen as an increase in velocity dispersion.  A typical value for the velocity dispersion of the clumpy disk gas is $\leq$10 km s$^{-1}$, while a higher-dispersion ($\sim$15 km s$^{-1}$) diffuse component is observed in some galaxies (e.g.\  \citealt{2013AJ....146..150C}, \citealt{2016AJ....151...34C}).  In Circinus, the velocity dispersion in the overdense region is 30--50 km s$^{-1}$, which is shown more quantitatively in Figure~\ref{dispersion_profile}. 

\par
   There is also an extension to the northwest (referred to as the ``NW cloud" in the text and circled in Figures).  The NW cloud is almost certainly associated with the outflow as it appears to be moving radially outward in projection in addition to being in apparent alignment with the ionized outflow cone.  Its velocity is also blue-shifted up to $\sim$150 km s$^{-1}$, similar to that of the ionized outflow. The overdense region and projected direction of the NW cloud coincide with the center of the outflow cone observed in the ionized gas (e.g.\ \citealt{1994Msngr..78...20M}, \citealt{1997ApJ...479L.105V}, \citealt{2010ApJ...711..818S}).

\begin{figure*}
\includegraphics[width=180mm]{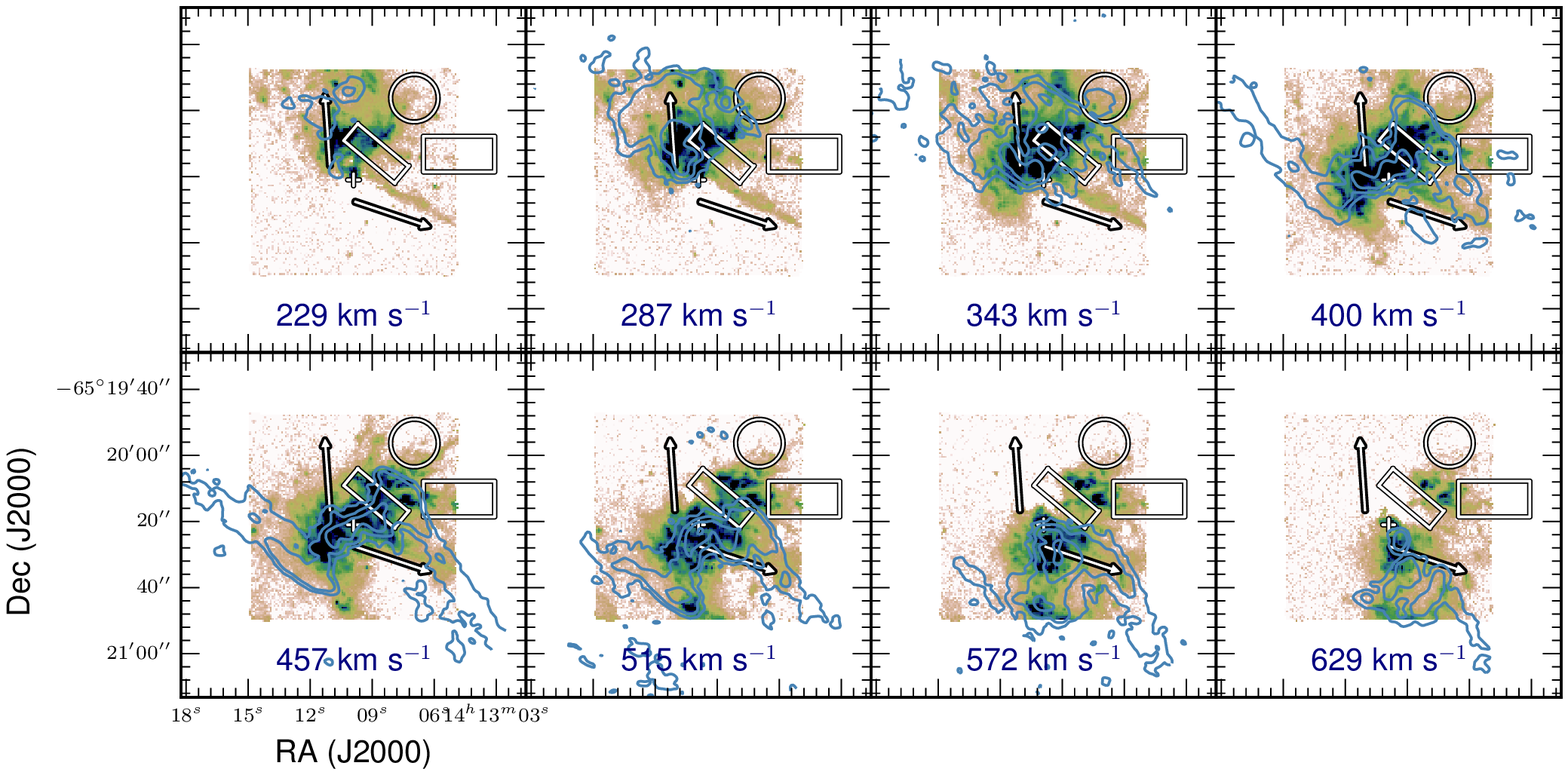}
\caption{\textit{Channel maps showing the H$\alpha$ as observed with MUSE (for a more complete presentation of those data, see Venturi at al. (\textit{in prep}) and Marconi et al. (\textit{in prep})) with CO (1-0) moment maps spanning the corresponding MUSE velocities overplotted in blue.  Contours begin at 7.5 mJy bm$^{-1}$ km s$^{-1}$) and increase by factors of four.  Marked features are as in Figure~\ref{halpha_muse}.  While the ionized outflow is prominently seen in straight filaments directed primarily to the northwest, its velocities coincide with those of the main disk.  The issue is compounded further for the molecular component as it also includes gas in the outer spiral arms.  The outer arms do not follow disk-rotation, thus rendering disentangling the molecular outflow from regions over which it intersects difficult.} \label{channel_maps_muse}}
\end{figure*}

\par
   There is a second extension to the west (referred to as ``far W cloud" in the text and is encompassed by the western rectangle in Figures).  Like the NW cloud, the far W cloud is beyond the northern spiral arm and appears to travel radially outward from Circinus.  The far W cloud also appears to extend along one of the H$\alpha$ filaments (Figure~\ref{channel_maps_muse}, 400 km s$^{-1}$).  Furthermore, the far W cloud and that the H$\alpha$ filament coincide in velocity.

\par
    Channel maps showing the NW and far west clouds are shown in Figure~\ref{channel_maps_clouds}.  In these channel maps, the NW cloud moves in a northwest direction with increasing velocity (inside the white circle).  While there is nearby emission from the northern spiral arm in these channels, it is essential to note that, although the morphology changes somewhat between channels, \textit{there is no such spatial shift in the northwest direction by the arm emission}.  Thus, while the NW cloud and the northern spiral arm coincide spatially in some channels (given the geometry, they inevitably intersect), their kinematics are distinct from each other.  The far W cloud moves almost directly westward (along the horizontal rectangle).  There is emission from the northern spiral arm to the southwest of the horizontal rectangle, but this emission does not appear to move westward as is the case with the western cloud.  Rather, the northern arm emission again remains in the same spatial location in each progressive channel. We emphasize that these effects are subtle, especially in the far W cloud. In the case of the far W cloud, additional observations at either higher resolution or more CO transitions are necessary to confirm whether it is indeed part of the outflow.

\section{Results}\label{results}

\subsection{The Molecular Outflow in Circinus}\label{wind}

\par
   Although we identify a region of higher density and increased velocity dispersion where a molecular outflow would be expected, the outflow cannot be easily separated from disk material.  Based on the ionized outflow (Figure~\ref{channel_maps_muse}), it is clear that much of the outflow overlaps and shares similar velocities to those of the disk in projection.  Even with the higher velocity resolution of ALMA (3 km s$^{-1}$) as opposed to MUSE (57 km s$^{-1}$), we find that the kinematics are still difficult to disentangle.  Furthermore, the orientation of the bar is such that it imposes higher surface-brightness and non-circular velocities near the base of the outflow.  The result is that the bar and outflow are indistinguishable (putting bar-symmetry arguments aside for now).  Finally, the outflow to the northwest intersects two spiral arms -- one within $\sim$7" and another at larger radii beyond the central region of the main disk.  The innermost spiral arm and bar are both avoided by not including the central 7.5" in our calculations.  The northern spiral arm covers a larger radial extent and shows velocities clearly distinct from the main disk.  Thus, we also exclude this overlapping region from our calculations.  With these exclusions, a limited region remains (white box in Figure~\ref{moments_CO}).  For our calculations, we assume that any anomalous or over-dense gas in this restricted region is indeed part of the molecular outflow.

\par
   To isolate potential outflowing material, we use four approaches:  The first approach involves integrating over regions in which outflowing material is expected based on the available multiwavelength data and then subtracting the flux from the main disk.  The second approach involves mirroring the velocity field of each half along the major axis and then subtracting them from each other in order to obtain a residual (e.g.\ \citealt{2016MNRAS.457.1257H} for ionized outflows). A third method involves creating kinematic models of the main disk and then subtracting them from the data in order to isolate the wind that remains in the residuals (e.g.\ \citealt{2014A&A...567A.125G}, \citealt{2016arXiv160703674P}).  The fourth approach considers a spectral analysis of select regions containing potential outflow features. The first two approaches are straightforward as presented in $\S$~\ref{flux} and~\ref{vel_resid}, while the third approach is more involved, thus we present the results from that method in $\S$~\ref{model_wind} describe that process in Appendix~\ref{modeling}.  The fourth method analyzes all potential outflow features that become evident via the first three methods, and is thus presented in $\S$~\ref{spectra}.

\subsubsection{The Molecular Outflow Based on Integrated Flux Measurements}\label{flux}

\par
 We consider emission in the overdense region only from approximately 7.5--15" (330--650 pc deprojected).  The region is enclosed by the white box in Figure~\ref{moments_CO}.  It is in-between spiral arms, which presents the possibility that some of the emission is from the spiral arms.  However, for this calculation, we assume that all emission in excess of the disk is due to an outflow.  To obtain this excess emission, we integrate the flux in the overdense region and do the same for the region of the disk to the southeast that is azimuthally opposite.  We then subtract the latter from the former.  The final value is the outflowing emission within the overdense region, which can then be converted into a luminosity in order to determine its molecular mass.

\subsubsection{Outflow Extraction via Inspection of Asymmetries in the Velocity Field}\label{vel_resid}

\par
   We employ a second method in which the velocity field is mirrored along the major axis and then subtracted in order to obtain a residual.  This approach is similar to that taken by \citet{2016MNRAS.457.1257H} for ionized outflows in edge-on galaxies.  While Circinus is moderately-inclined, the same principles apply to first order. If a large difference is seen in the residuals outside of the effective radius of the molecular disk, and thus likely to be extra-planar, then it is an indicator of outflowing gas. While there are some differences as seen in Figure~\ref{mirror_residual}, the molecular gas is clearly not outflow-dominated.  The greatest extremes are in the outermost spiral arms, but these are due to motions within the spiral arms themselves and not associated with a large-scale outflow.

\begin{figure*}
\includegraphics[width=180mm]{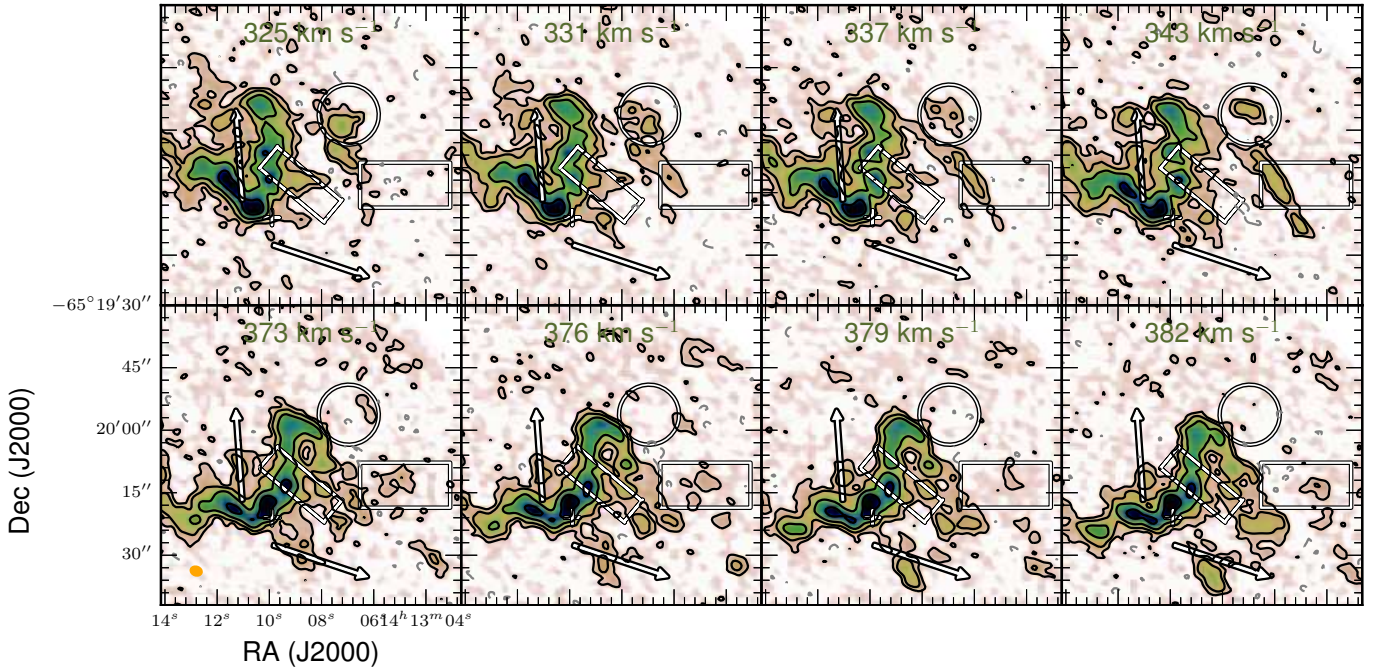}
\caption{\textit{Channel maps showing the NW cloud (top row) and the far W cloud (bottom row).   Contours are as in Figure~\ref{channel_maps} and marked features are as in Figure~\ref{halpha_muse}.  The NW cloud moves in a northwest direction with velocity while the far W cloud moves in a western direction.} \label{channel_maps_clouds}}
\end{figure*}

\subsubsection{Extracting the Outflow Through Kinematic Modeling}\label{model_wind}
    
\par
  We utilize kinematic modeling as detailed in Appendix~\ref{modeling} to further separate the molecular outflow from the main disk, as well as the spiral arms, which presumably share the circular motions of the main disk.  Residual moment maps are shown in Figure~\ref{residual_moments_all}.   The first set is after subtracting a simple disk model with no additional features.  A second set of residuals is created by subtracting a simple disk, bar, and spiral arms (with radial motions).  

\par
     The overdense region still contains substantial emission after the disk-only subtraction, while there is none remaining in the azimuthal range 180$^{\circ}$ to the southeast.    The flux remaining in the overdense region has clearly distinct kinematics from the nearest remaining flux associated with the northern spiral arm.   Curiously, this remaining emission in the overdense region has similar kinematics to those of the northwestern cloud, which is almost certain to be outflowing molecular gas.  The remaining emission is also aligned in the same direction away from the disk as is the NW cloud.

\par
    Additional molecular gas is eliminated in the residuals created by subtracting models containing additional features.  Much of the emission in the overdense region could alternatively be attributed to the spiral arm.  However, the blue-shifted feature remains and is still aligned with the NW cloud.  

\par
    The far W cloud also remains in the final residuals, but it is clear that some of the gas on its eastern edge is indisputably associated with the northern spiral arm.  However, this remaining gas from the arm is unsurprising given its presence in the panel corresponding to 340 km s$^{-1}$ in Figure~\ref{channel_maps} -- the edges of the far W cloud and northern spiral arm simply overlap spatially, but at different velocities.  It is demonstrated in Appendix~\ref{modeling} that not all of the northern spiral arm emission is eliminated in the residuals due to the difficulty in fully-constraining the streaming motions within them.

\par
   Another feature, starting from the nucleus and extending almost directly north (along the northern arrow), becomes clear.  This feature (henceforth referred to as the ``northern stripe") can also be seen in channel maps (Figures~\ref{channel_maps} and Figure~\ref{channel_maps_clouds}) extending from 340--460 km s$^{-1}$ in the former and all channels in the latter.  In some regions along the extent if the northern stripe, it is blue-shifted as much as 110 km s$^{-1}$ with respect to the peak (at the same pixel location), but blue-shifted by approximately the same amount from systemic.  The northern stripe is near the edge, but still within the extent of the wide-angled ionized outflow cone.  The relatively extreme location of the northern stripe when compared to the central filaments renders it much simpler to disentangle than any other possible outflow components.  Thus, a scenario in which only a single filamentary structure associated with the outflow near the edge of the cone is detected is possible.

\par
    Detailed descriptions of the models are given in Appendix~\ref{modeling}.

\begin{figure}
\begin{centering}
\includegraphics[width=80mm]{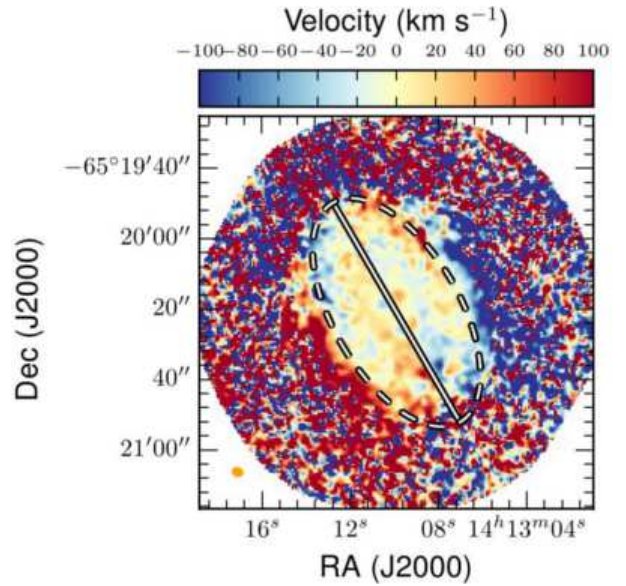}
\caption{\textit{The residual resulting from subtracting southeastern the velocity field from the northwestern velocity field mirrored across the major axis.  The line in the center is directly along the major axis, while the ellipse represents the radius of the projected molecular disk (not including the outermost spiral arms).  The beam is shown in the lower left corner.  If the disk were outflow-dominated, there would be a substantial difference in the two halves (appearing as red and blue extremes in color and dark regions in grayscale).  The most extreme velocity differences are in the outermost spiral arms just beyond the ellipse.  These extremes are due to motions within the spiral arms rather than related to an outflow from the nuclear regions.} \label{mirror_residual}}
\end{centering}
\end{figure}

\begin{figure*}
\begin{centering}
\includegraphics[width=160mm]{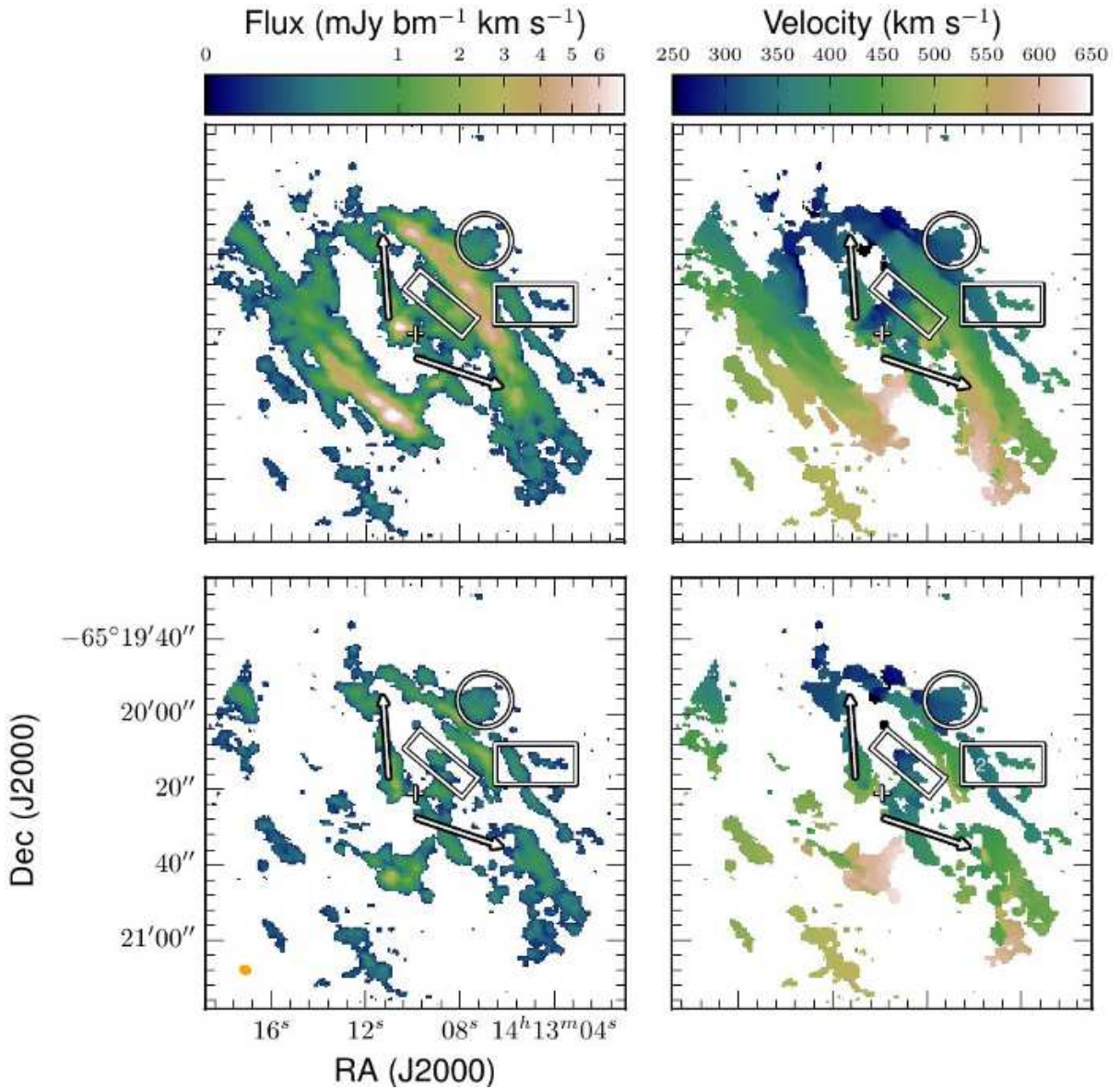}
\caption{\textit{Zeroth- (left) and first- (right) moment maps created from two residuals clipped with a 30 mJy bm$^{-1}$ threshold for clarity. The residuals after subtracting a simple disk model with no additional features are shown in the top row and residuals after subtracting a simple disk, bar, and spiral arms (with radial motions) are shown in the bottom row. Outlined regions are as in Figure~\ref{halpha_muse}.  The beam is shown in the lower, left corner in orange.   The overdense region still contains substantial emission after the disk-only subtraction.  The remaining flux in the overdense region after subtracting the final models (bottom, right panel) has clearly distinct kinematics (blue-shifted $\sim$150 km s$^{-1}$) from the nearest remaining flux associated with the northern spiral arm in the same panel (denoted by ``1").   This remaining emission in the overdense region has similar kinematics to the northwestern cloud - an almost certain outflow feature. Kinematically distinct sections of the spiral arms remaining after the final models are denoted by ``1" and ``2" to emphasize the distinction.  The blue-shifted emission directly south of the overdense region corresponds to the increased velocity dispersion to the south first introduced in Figure~\ref{moments_CO}.} \label{residual_moments_all}}
\end{centering}  
\end{figure*}

\subsubsection{A Spectral Analysis of the Outflow}\label{spectra}

\par
\begin{figure}
\begin{centering}
\includegraphics[width=80mm]{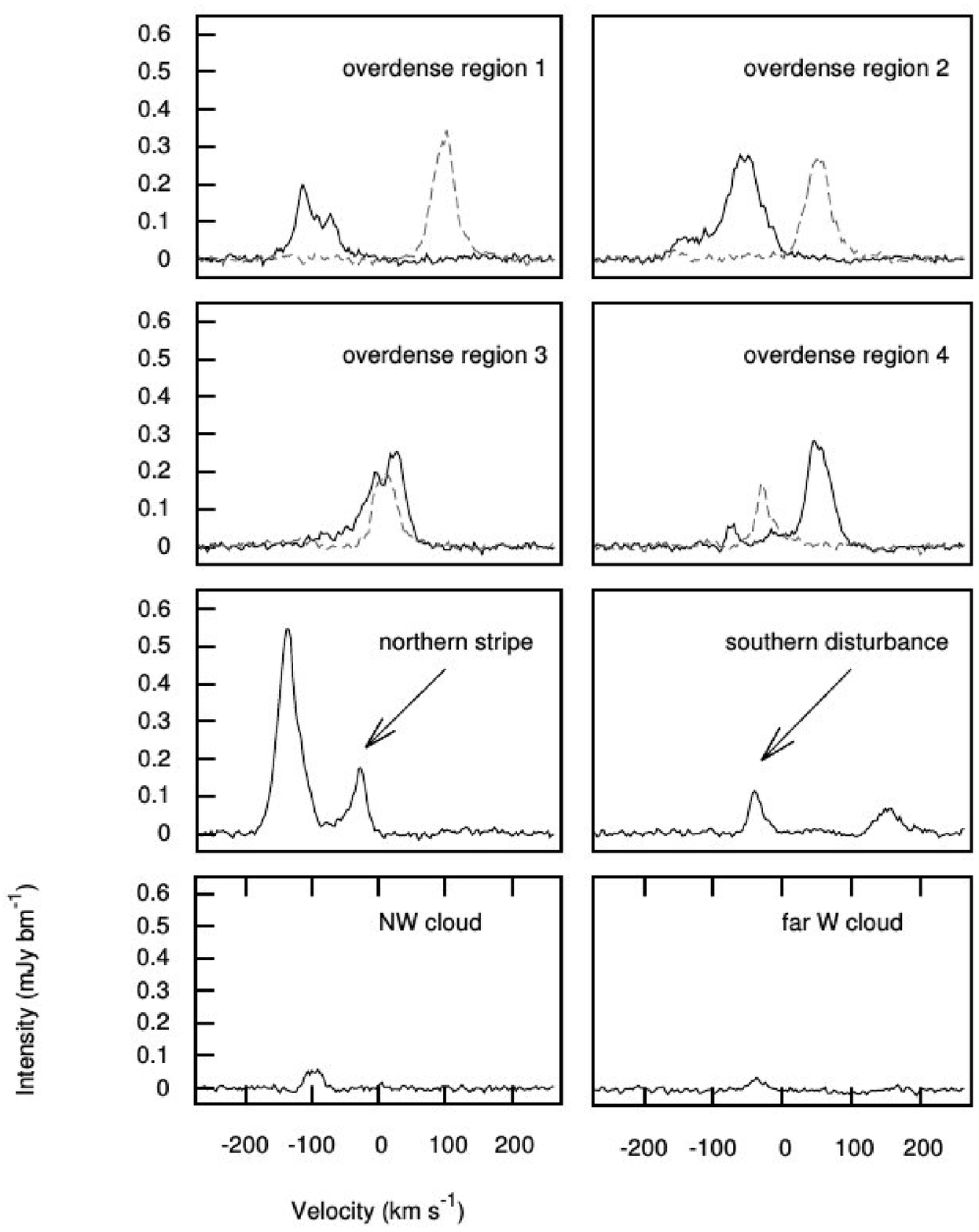}
\caption{\textit{Representative spectra from regions containing molecular gas with anomalous velocities.  Spectra locations are shown in Figure~\ref{spectra_map} and correspond to the features originally shown in Figure~\ref{halpha_muse}.  The top four panels show spectra from the overdense region (black), with the corresponding spectra that are azimuthally opposite (gray, dashed).  With the exception of the top, left panel (corresponding to the northernmost overdense region spectrum and the southernmost azimuthally opposite spectrum), all of the spectra in the overdense region have low-level blue-shifted wings.  Spectra in the azimuthally opposite region show hints of red-shifted wings.} \label{spectra_grid}}
\end{centering}
\end{figure}
\begin{figure}
\begin{centering}
\includegraphics[width=80mm]{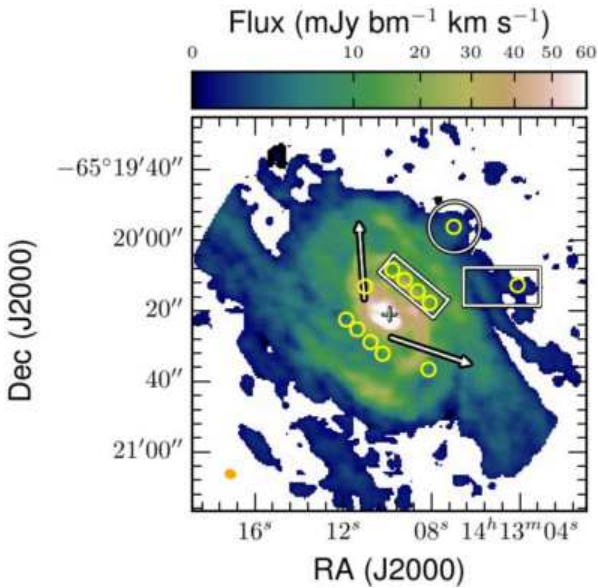}
\caption{\textit{A zeroth-moment map showing the locations of spectra in Figure~\ref{spectra_grid}.  The spectra are single pixels at the center of each yellow circle.  The northernmost circle in the overdense region represents ``overdense region 1" in Figure~\ref{spectra_grid}, going in order to ``overdense region 4" to the southwest.  Spectra in the azimuthally opposite region are in reverse order, starting from the southernmost (``region 1") and going to the northeast.} \label{spectra_map}}
\end{centering}
\end{figure}

    Now that all anomalous features have been identified, we present representative spectra of these regions in Figure~\ref{spectra_grid}, with the spectra locations shown in Figure~\ref{spectra_map}.  The spectra in the overdense region consistently show blue-shifted wings.  The northern stripe is blue-shifted from systemic, but red-shifted with respect to the approaching half of the galaxy.  The former velocity shift is relevant when considering the outflow.  The southern disturbance is again clearly seen as blue-shifted from systemic.  The NW and far W clouds are both blue-shifted ($\sim$100 km s$^{-1}$ for the former and 40--50 km s$^{-1}$ for the latter).  

\par
\begin{figure}
\begin{centering}
\includegraphics[width=80mm]{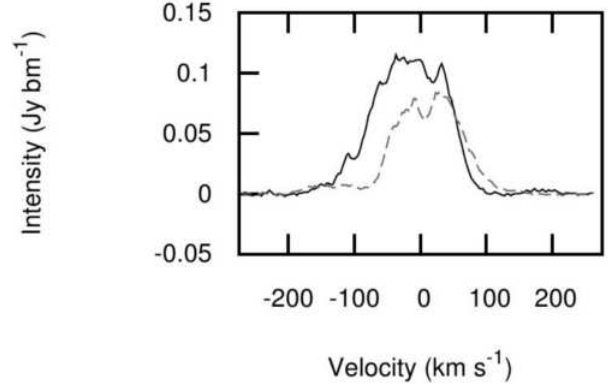}
\caption{\textit{Average spectrum over the overdense region (black, solid) and the region azimuthally opposite (gray, dashed).} \label{spectra_average_overdense}}
\end{centering}
\end{figure}

   An average spectra of the overdense region and azimuthally opposite region is shown in Figure~\ref{spectra_average_overdense}.  The blue-shifted wing remains in the average spectrum for the overdense region, but there are no indications of a similar blue-shifted feature in the region that is azimuthally opposte.  There are, however, slight indications of a red-shifted wing.

\subsubsection{Molecular Outflow Parameters Determined from the Data}\label{numbers}

\par
The first method considers the excess flux present in the overdense region when compared to the azimuthally opposte region to the southeast.  This method provides a basic estimate for the mass of outflowing molecular gas.  The second method provides no quantifiable results, but is rather a confirmation that Circinus is not outflow-dominated.  The third method can, in theory, yield a mass estimate for the outflow, but not all of the emission can be disentangled from the main disk due to the shared rotation of the disk and outflow.  All but $\sim$25$\%$ of the potential outflowing material associated with the overdense region (first method) is eliminated by a disk-only subtraction, and all but a few percent is eliminated by our final model, which includes a bar and spiral arms (features with kinematic properties that cannot be unambiguously constrained).  Because of the shared outflow/disk rotation that is known to exist in the ionized component, it is obvious that most of the potential outflowing molecular gas will be subtracted by this method and such a result should not be considered valid.  Ultimately, a reliable outflow mass cannot be derived from this method in the case of Circinus.  While the fourth method reinforces evidence for a molecular outflow, the spectra do not provide an improved means for determining the molecular mass itself.  However, the fourth method provides clearer velocity measurements for the potential outflowing components. For these reasons, we use the numerical values derived using the first method for the mass-related quantities, and velocities derived from the fourth method (or optical in the case of the overdense region and upper limits) in the following calculations.

\par
The CO luminosity associated with the outflow in the overdense region is 3.8$\times$10$^{6}$ K km s$^{-1}$ pc$^{2}$.  Using the $\alpha_{CO}$ conversion factor of 0.34 M$_{\odot}$ pc$^{-2}$ (K km s$^{-1}$)$^{-1}$ put forth in \citet{2013ARA&A..51..207B} for the lower limit of outflows, we obtain a molecular mass of 1.6$\times$10$^{6}$ M$_{\odot}$, which is on the order of 0.3--0.5$\%$ of the total molecular mass of the disk out to 2.6 kpc.    Another reason to consider this $\alpha_{CO}$ value a lower limit is that \citet{2015ApJ...814...83L} find that their observed low-J CO line-ratios are difficult to reproduce in the optically thin case in the outflow of M 82, thus indicating that it is not necessarily a valid assumption for the outflowing emission.   

\par
    In addition to the overdense region, we include the NW and far W clouds, as well as the northern stripe in our calculations.  The southern disturbance is not included as it shows strong indications of being associated with the northern spiral arm, especially when considering its furthest extent to the southwest.

\par
We determine the CO luminosity of the NW cloud as 6.2$\times$10$^{5}$ K km s$^{-1}$ pc$^{2}$ and its molecular mass to be 2.1$\times$10$^{5}$ M$_{\odot}$ using $\alpha_{CO}$=0.34 M$_{\odot}$ pc$^{-2}$ (K km s$^{-1}$)$^{-1}$.  The values for the far W cloud under the same assumptions are 9.2$\times$10$^{4}$ K km s$^{-1}$ pc$^{2}$ and 3.1$\times$10$^{4}$ M$_{\odot}$.  Finally, the corresponding values for the northern stripe are 1.2$\times$10$^{6}$ K km s$^{-1}$ pc$^{2}$ and 4.2$\times$10$^{5}$ M$_{\odot}$.   The luminosities for the overdense region and far W cloud have no measured uncertainties as the boundaries for these features are clear.  In the case of the overdense region, the box is clearly defined.  The measurement of the far W cloud does not include any ambiguous diffuse emission or overlap with any other structures.  The NW cloud and the northern stripe, however, have measured uncertainties in their CO luminosities due to their proximity to larger-scale structures.  These uncertainties are on the order of 30$\%$.

\par
   We now determine the mass outflow rate.  As was also done in \citet{2013Natur.499..450B} for the case of NGC 253, we use the following formula based on the morphology of the outflow:
\begin{equation}
\dot{M}=\frac{2\alpha_{CO}L_{CO}v_{w}\tan{i}}{r_{w}}\label{equation_outflow}
\end{equation}
where $L_{CO}$ is the CO luminosity (already multiplied by $\alpha_{CO}$, the conversion factor discussed in the last section), $v_{w}$ is the projected velocity of the outflow, $i$ is the inclination of the outflow, and $r_{w}$ is the extent of the outflow in projection.  For an initial estimate, we assume $v_{w}=150$ km s$^{-1}$ based on the highest velocities we see in the molecular component and the optical velocities (180 km s$^{-1}$) as seen in \citet{1997ApJ...479L.105V}.  Based on the NW cloud, the outflow is estimated to extend $\sim$35" (700 pc in projection, 1.5 kpc deprojected).  However, the overdense region over which we integrate only extends from 7.5" (150 pc in projection, 0.3 kpc deprojected) to 22.5" (450 pc in projection, 1.0 kpc deprojected).  Thus, for calculating the outflow rate of gas associated with the overdense region, we assume this range.  While the outflow cone must have some range of inclinations with respect to the observer, for simplicity we assume it is perpendicular to the disk inclination of 65$^{\circ}$.  Based on these assumptions, we determine a molecular mass outflow rate of 3.0 M$_{\odot}$ yr$^{-1}$ in the overdense region.  While we use lower limit assumptions for this calculation, this value should not be considered a lower limit in the strictest sense as there may be material included in our mass estimate that is actually part of the spiral arms. Rather, it is a lower limit provided that the excess flux in the overdense region is indeed due to outflowing material.

\par
    To derive an upper limit for the overdense region, we assume a maximum velocity of 180 km s$^{-1}$ as in \citet{1997ApJ...479L.105V}. We also assume a higher $\alpha_{CO}$ of 0.8 M$_{\odot}$ pc$^{-2}$ (K km s$^{-1}$)$^{-1}$ for an optically-thick case.  These assumptions yield a mass outflow rate of 8.3 M$_{\odot}$ yr$^{-1}$ in the overdense region.  

\par
    The NW cloud also contributes to the molecular mass outflow rate.  Assuming the cloud extends from roughly 25--35" (510--700 pc in projection, 1.1--1.5 kpc deprojected), a velocity of 100 km s$^{-1}$, and $\alpha_{CO}$ of 0.34, we obtain a lower limit of 0.35 M$_{\odot}$ yr$^{-1}$.  In contrast to the value calculated for the overdense region, \textit{this is a strict lower limit} because the NW cloud appears to be outflowing from the galaxy.  Assuming 180 km s$^{-1}$ (on the higher-end of the optical velocities), and $\alpha_{CO}$ of 0.8 an upper limit for the molecular mass outflow rate of the NW cloud is 2.5 M$_{\odot}$ yr$^{-1}$.  These values also consider the $\sim$30$\%$ uncertainty in the CO luminosity measurement. 

\par
    The far W cloud contributes a negligible amount to the molecular mass outflow rate.  We use the same approximations as for the NW cloud, with the exception of the velocity, which we assume to be 50 km s$^{-1}$, and obtain a lower limit of 0.03 M$_{\odot}$ yr$^{-1}$ and an upper limit of 0.08 M$_{\odot}$ yr$^{-1}$.

\par
    Also assuming 50 km s$^{-1}$, but a total radial extent of 20", we find the molecular mass outflow rate associated with the northern stripe is 0.2--0.7 M$_{\odot}$ yr$^{-1}$.   These values also consider the $\sim$30$\%$ uncertainty in the CO luminosity measurement.  

\par
    If one includes the overdense region, clouds, and northern stripe in the total, the molecular mass outflow rate is 3.6--12.3 M$_{\odot}$ yr$^{-1}$.  We again emphasize that the lower limit assumes that all excess flux in the overdense region is due to outflowing molecular gas.  The strictest lower limit is 0.35 M$_{\odot}$ yr$^{-1}$, which only includes the outflowing NW cloud.

\par
  \citet{2012MNRAS.425.1934F} list a range of possible global star formation rates for Circinus, derived using a variety of methods by different authors.  We adopt the value of 4.7 M$_{\odot}$ yr$^{-1}$ calculated by \citet{2008ApJ...686..155Z} using the total infrared luminosity (L$_{TIR}$). The molecular mass outflow rate of 0.35--12.3 M$_{\odot}$ yr $^{-1}$ is a comparable to the global star formation rate in Circinus and thus may significantly affect star formation in its central regions.

\par
    We find the molecular mass of the observed disk (out to $\sim$1.5 kpc) to be 2.9$\times$10$^{8}$$\pm$8.7$\times$10$^{7}$ M$_{\odot}$ assuming an $\alpha_{CO}$ of 4.3 M$_{\odot}$ pc$^{-2}$ (K km s$^{-1}$)$^{-1}$ from \citet{2013ARA&A..51..207B}.  This $\alpha_{CO}$ has an uncertainty of $\sim$30$\%$ and is based on typical values for the Milky Way disk that are derived using a variety of methods, including CO isotopologues, and dust extinction.  This molecular disk mass indicates that if the gas were actually leaving the galaxy with a constant mass outflow rate of 0.35--12.3 M$_{\odot}$ yr $^{-1}$, the molecular gas would be depleted on the order of 10$^{7}$--10$^{8}$ yr.  Although, with the outflow velocity being substantially less than the escape velocity, the molecular gas will most likely be reaccreted. This conclusion is based on the peak rotational velocity of $\sim$200 km s$^{-1}$ (derived in Appendix~\ref{modeling}), which yields an escape velocity $\geq$500 km s$^{-1}$ near the center \citep{2008gady.book.....B}.  Alternatively, \citet{2005ApJ...621..227M} provides an estimate on the escape speed from the halo of a few times the rotational velocity.  The projected velocity of the outflow does not come close to reaching either estimate (although we cannot be certain about velocities across the line of sight).

\par
   The kinetic power ($P_{kin, OF}=\frac{1}{2}v^2\dot{M}_{H_{2}, OF}$) and momentum rate ($v\dot{M}_{H_{2}, OF}$) of the molecular outflow can also be estimated from these data.  Using the lower and upper limits for the mass outflow rates determined above, we find a range of 1.0$\times$10$^{39}$--1.1$\times$10$^{41}$ erg s$^{-1}$ for the $P_{kin, OF}$, and 2.0$\times$10$^{32}$--1.3$\times$10$^{34}$ g cm s$^{-2}$ for ($v\dot{M}_{H_{2}, OF}$).

\subsection{Additional Detected Lines}

\par
   In addition to CO(1--0), we detect two groups of CN (1--0) radicals - four hyperfine J=1/2 transitions at $\sim$ 113.15 GHz, and five J=3/2 hyperfine transitions at $\sim$113.5 GHz.  While the fine structure of the lines is difficult to disentangle from each other, the rest frequencies of each transition are 113.191--113.123 GHz in the former group and 113.488-113.520 GHz in the latter.  CN is a photodissociation region (PDR) tracer (e.g.\ \citealt{1985ApJ...291..722T}), which is consistent with it being heavily-concentrated at the nucleus as we see in Circinus.  

\par
   C$^{17}$O is also detected in Circinus -- one of very few extragalactic detections to date (e.g.\ NGC 253, M82 and IC 342 by \citealt{1991A&A...249...31S}, M 51 by \citealt{2014ApJ...788....4W}).  C$^{17}$O is a secondary nucleosynthesis product of low/intermediate mass star formation.  The current data do not provide detailed maps of its distribution. However, as suggested by \citet{1991A&A...249...31S}, by mapping its distibution, deeper and higher resolution observations will yield insight concerning where different mass star formation is taking place (e.g.\ when compared with CO (1--0), a high ratio of C$^{17}$O indicates low/intermediate star formation, a low ratio indicates massive star formation).

\begin{deluxetable*}{lcc}
\tabletypesize{\scriptsize}
\tablecaption{Additional Molecular Species \label{tbl_2}}
\tablewidth{0pt}
\tablehead
{ 
\colhead{Observed Frequency (GHz)}&
\colhead{Molecular Species} &
\colhead{Rest Frequency (GHz)}
}
\startdata
\\

\phd 112.95\tablenotemark{a}&CN(1--0), J=1/2--1/2, F=1/2--1/2&113.123 \\ 
\phd 112.98&\hspace{80pt} F=1/2--3/2&113.144 \\
\phd 113.00&\hspace{80pt} F=3/2--1/2&113.170 \\
\phd 113.01&\hspace{80pt} F=3/2--3/2&113.191 \\
\phd 113.30--113.33&CN(1--0), J=3/2--1/2, F=3/2--1/2&113.488 \\
\phd 113.30--113.33&\hspace{80pt} F=5/2--3/2&113.491 \\
\phd 113.30--113.33&\hspace{80pt} F=1/2--1/2&113.500 \\
\phd 113.30--113.33&\hspace{80pt} F=3/2--3/2&113.509 \\
\phd 113.30--113.33&\hspace{80pt} F=1/2--3/2&113.520 \\
\phd 112.18&C$^{17}$O &112.359 \\
\tablenotetext{a}{Taken from peak central position.}

\enddata
\end{deluxetable*}

\section{Discussion}\label{discussion}

\subsection{Energetics of the Outflow}

\par
   We now consider how the AGN and star formation in Circinus may be affecting the outflow properties.

\par
   Circinus hosts a 10$^{10}$ L$_{\odot}$ AGN \citep{1996A&A...315L.109M} and has a bolometric luminosity (L$_{bol}$) of 1.7$\times$10$^{10}$ L$_{\odot}$ (\citealt{1997A&A...325..450S}, \citealt{1998ApJ...493..650M}).  These values yield $L_{AGN}$/$L_{bol}$ of around 0.6 (closer to 0.9 if using values derived in \citealt{2016ApJ...826..111S}).  L$_{AGN}$ for Circinus is comparable to that of NGC 1266, NGC 1068, and NGC 1377 (\citealt{2014A&A...562A..21C} and references therein), and thus on the lower-end of AGN luminosities for the molecular outflows observed to date.  $L_{AGN}$/$L_{bol}$ for Circinus, however, is 2--3 times higher than for those galaxies.

\par
   Relevant to the launching mechanism of the outflow are its kinetic power and momentum rate.  Observations show that these properties, when associated with AGN-driven outflows, show some correlation with L$_{AGN}$ (e.g.\ \citealt{2014A&A...562A..21C}, \citealt{2015A&A...583A..99F}).  Such connections have also resulted from simulations (e.g.\ \citealt{2012ApJ...745L..34Z}, \citealt{2012MNRAS.425..605F}).

\par
\begin{figure*}
\begin{centering}
\includegraphics[width=180mm]{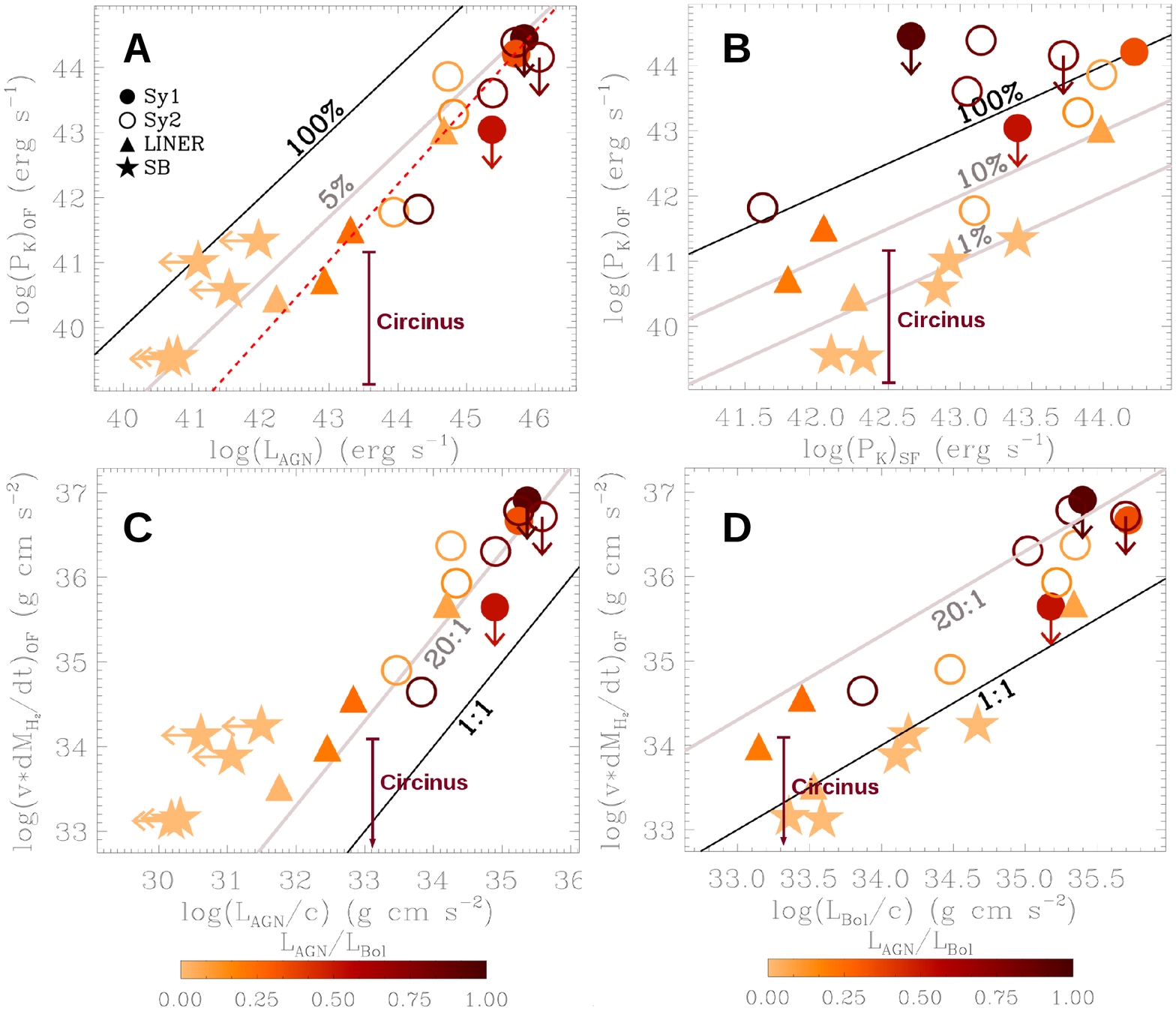}
\caption{\textit{Figures 12 (panel A), 13 (panel B), 14 (panel C), and 16 (panel D) reproduced directly from \citet{2014A&A...562A..21C} (with permission), but with the addition of Circinus here. Symbols labeled in panel A correspond to Seyfert 1 (Sy1), Seyfert 2 (Sy2), LINER and starburst (SB) galaxies.  The ranges for P$_{kin,OF}$ and $v\dot{M}_{H_{2}, OF}$ for Circinus have been added where applicable.  It is important to note that $\alpha_{CO}$=0.8 was assumed for the galaxies in the \citet{2014A&A...562A..21C} work (upper end of range shown for Circinus).  In all cases, P$_{kin,OF}$ and $v\dot{M}_{H_{2}, OF}$ for Circinus is significantly lower than that of the AGN-driven outflows, especially those with high $L_{AGN}$/$L_{bol}$.  However, when considering the upper-limit of $\alpha_{CO}$=0.8 (typically used for ULIRGs) in the case of Circinus, it appears to follow the relation indicated by the Seyfert 2 galaxies when extrapolated to lower $L_{AGN}$.} \label{cicone_all_mod}}
\end{centering}
\end{figure*}

   The kinetic power of 1.0$\times$10$^{39}$--1.1$\times$10$^{41}$ erg s$^{-1}$ that we determine for the molecular outflow is generally lower than that of the \citet{2014A&A...562A..21C} sample, especially the outflows assumed to be driven by AGN.  In that work, a relation was established between the kinetic power of the outflows and the L$_{AGN}$ of the host galaxies (generally finding that the P$_{kin,OF}$ is $\sim$5$\%$ of L$_{AGN}$).  Circinus has a substantially lower P$_{kin,OF}$/L$_{AGN}$ ratio (on the order of 0.01--0.3$\%$ of L$_{AGN}$) compared to the ULIRGs and other Seyferts.   However, this range is consistent with the linear fit expressed in Equation 3/Figure 12 of \citet{2014A&A...562A..21C}.   We directly duplicate Figure 12 from that work here with the addition of Circinus (Figure~\ref{cicone_all_mod}, panel A).  In that relation, P$_{kin,OF}$/L$_{AGN}$ decreases for lower L$_{AGN}$, indicating that the trends derived from that sample may indeed apply to lower luminosity galaxies as well.  The relation does not, however, appear to be consistent with the high ratio of L$_{AGN}$ to L$_{bol}$ in Circinus.  

\par
   The kinetic power due to supernovae (P$_{kin,SF}$) is also useful to consider.  Equation 2 from \citet{2005ARA&A..43..769V} indicates that P$_{kin,SF}$=7$\times$10$^{41}$ SFR(M$_{\odot}$ yr$^{-1}$), yielding a value of 3.5$\times$10$^{42}$ erg s$^{-1}$ for Circinus -- again low given its AGN properties.   When P$_{kin,OF}$ is plotted against P$_{kin,SF}$, P$_{kin,OF}$ for Circinus is well below that of the AGN-driven outflows with $L_{AGN}$/$L_{bol}$ summarized in Figure 13 of \citet{2014A&A...562A..21C}, which we show here in panel B of Figure~\ref{cicone_all_mod}.  Instead, Circinus falls within the LINER and starburst population.

\par
    We find the momentum rate of the molecular outflow is 2.0$\times$10$^{32}$ --1.3$\times$10$^{34}$ g cm s$^{-2}$.  With these values, we find a ratio of  $v\dot{M}_{H_{2}, OF}$ to L$_{AGN}$/$c$ of 0.1--10.3.  \citet{2014A&A...562A..21C} found this ratio to be $\sim$20 (and often significantly higher) for outflows in host galaxies having L$_{AGN}{\geq}$10$\%$ of L$_{bol}$.  The ratio found in that work is consistent with simulations presented by \citet{2012ApJ...745L..34Z}.  Circinus is well below this ratio, again indicating that it does not have similar outflow properties to other galaxies with high values of $L_{AGN}$/$L_{bol}$.  However, if one considers only the ratio of $v\dot{M}_{H_{2}, OF}$/L$_{AGN}$, the upper-limit for Circinus derived using $\alpha_{CO}$=0.8 (typically used for ULIRGs) is again consistent with the \citet{2014A&A...562A..21C} trend seen in the galaxies with presumed AGN-driven outflows if one extrapolates to lower L$_{AGN}$ (Figure 14 of \citealt{2014A&A...562A..21C} or panel C of Figure~\ref{cicone_all_mod} here).  When $v\dot{M}_{H_{2}, OF}$ is considered only in relation to $L_{bol}$, Circinus again appears to be consistent with starburst galaxies and LINERS (Figure 16 of \citealt{2014A&A...562A..21C} or panel D of Figure~\ref{cicone_all_mod} shown here).

\par
   In summary, when considering the kinetic power and momentum rate, Circinus appears to be consistent with trends put forth in \citet{2014A&A...562A..21C}, with the exception of those connected to its high AGN luminosity (and $L_{AGN}$/$L_{bol}$).

\subsection{The Multiphase Structure of the Wind}

\par
    While there is a clear ionized outflow in Circinus, the existence of one in molecular phase is ambiguous. In the preceding sections we have presented the strongest evidence for one within the ALMA data:  1)  There is an overdensity to the northwest of the nucleus.  This overdense region coincides with the location of the ionized outflow cone.  2)  There is a cloud to the northwest (NW cloud) that appears to be traveling away from the disk.  The NW cloud is also blue-shifted by approximately the same amount as the ionized outflow. This feature is the clearest outflow signature in the ALMA data.   Based on the NW cloud, the molecular outflow is driven $\sim$35" in projection from the center, or 1.5 kpc when deprojected.  3) A similar, but less prominent cloud is seen to the west (far W cloud).  The far W cloud is directly along one of the ionized filaments, with velocities that coincide with that filament. 

\par
    The ionized outflow to the northwest in Circinus is well-known, manifesting as a $\sim$1.5 kpc blue-shifted outflow cone (e.g.\ \citealt{1994Msngr..78...20M}, \citealt{1997ApJ...479L.105V}, \citealt{2000AJ....120.1325W}, \citealt{2010ApJ...711..818S}).  All of these studies have noted the wide opening angle of the outflow cone, as well as filamentary structures.  \citealt{1997ApJ...479L.105V} note the presence of bow-shocked features, indicating strong interactions with the ISM.  \citealt{2010ApJ...711..818S} found evidence for a filled ionized gas cone and velocity gradients along filaments.  The current resolution of the molecular data does not allow for a direct comparison with the ionized outflow, but the features that we see have similar velocities. Like the ionized component, it is possible that there exists unidentifiable molecular outflow material is kinematically indistinct from disk emission.  

\par
   While we detect outflow signatures only on the near-side of the galaxy (also consistent with ionized observations), a counter-jet indeed exists as shown by \citet{1998MNRAS.297.1202E} in the radio continuum. If the aformentined overdense region is truly due to a molecular outflow, the lack of a molecular component to the counter-jet is somewhat intriguing as it was previously thought that the lack of a symmetric ionized component was due to absorption.  The counter-jet presented in \citet{1998MNRAS.297.1202E}, however, shows a substantial gap ($\sim$2' or 5--6 kpc taking projection effects into account) between the main disk and the radio emission on that side, indicating that perhaps any molecular component is either weaker on that side, or no longer present. There is also the possibility of asymmetries in the disk that could hinder the expulsion of material on one side.

\par
   Interestingly, \citet{2016ApJ...826..111S} search for OH outflows in more than 50 low-luminosity Burst Alert Telescope (BAT) detected AGN, including Circinus.  In the case of Circinus, they find an unambigous OH \textit{inflow} in the form of an inverted P-Cygni profile within the innermost $\sim$10" from the center.  Most of this region is omitted from our analysis due to resolution limitations.  Thus, a valid comparison must wait for higher resolution data.

\subsection{Implications for the Evolution of Circinus}

\par
   As demonstrated, it is difficult to detect outflowing molecular gas in Circinus. The kinematics of the disk and any potential outflow are largely overlapping as demonstrated in Figure~\ref{channel_maps_muse} for the ionized outflow. We see multiple indications of a molecular outflow in the ALMA data, including the overdense region, the NW cloud and far W clouds, and the northern stripe.  Still, there is no indication in the molecular phase of a prominent outflow such as what is seen in the ionized component (which is clearly seen even without kinematic information).  Furthermore, even when we consider all potential outflowing molecular gas, its mass, mass outflow rate, and energetics are substantially lower than what is seen in AGN-driven outflows \citep{2014A&A...562A..21C}, especially when considering its high L$_{AGN}$/L$_{bol}$ ratio.  We now consider possible explanations for this result.

\par
     Circinus has one of the most highly-ionized narrow-line regions, which is visible in the optical, near-infrared, and mid-infrared (e.g.\ \citealt{1998ApJ...505..621C}, \citealt{2010MNRAS.402..724P}).  \citealt{1997ApJ...479L.105V} find evidence that the higher-resolution ionized component is launched from the region containing AGN.   The ionized component of the outflow outflow from the central regions of Circinus is prominent (e.g.\ \citealt{1994Msngr..78...20M}, \citealt{1997ApJ...479L.105V}, \citealt{2010ApJ...711..818S}).  Thus, it is likely that a majority of the gas in the outflow cone of Circinus is in the ionized phase, resulting in a weak detection of outflowing CO.

\par
   It is difficult to say whether the molecular component of the outflow is entrained gas, or if the ionized component condenses and forms molecular gas in-situ.  Higher resolution observations (especially on a level comparable to the MUSE data) and additional CO transitions, as well as shock tracers will help to answer this question.

\section{Summary}\label{summary}

\par
1)  We observe the central $\sim$1' of Circinus galaxy with ALMA, targeting the CO(1--0) emission line, also detecting CN (1--0) (J=3/2--1/2 and F=3/2--1/2) and C$^{17}$O.
\par
2)  The molecular disk exhibits primarily circular rotation and shows indications of a bar out to $\sim$10" (220 pc), as well as spiral arms.  The spiral arms exhibit mild asymmetries and extend beyond the rest of the molecular disk in radius.  The spiral arms also appear to have prominent (over 100 km s$^{-1}$) streaming motions.
\par
3)  There exists an overdensity of molecular gas to the northwest of the center of the galaxy that does not appear to be associated with the spiral arms (although it likely intersects them, especially at larger radii).  The overdense region also has a higher average velocity dispersion than the rest of the disk, and shows indications of non-circular motions.  The gas in this region appears to be blue-shifted compared to the main disk, consistent with the velocities of the ionized outflow.  Most of the anomalous gas in this region is however, faint compared to the rest of the disk, indicative of a diffuse outflow. 
\par
4)  Based on residuals after removing gas having circular motions, we see some indication of filamentary structures entrained in the wind.  These structures require higher resolution and multiphase data to confirm.  
\par
5)  The total molecular mass of the disk within $\sim$1.5 kpc is 2.9$\times$10$^{8}$$\pm$8.7$\times$10$^{7}$ M$_{\odot}$.  With the same assumptions, the molecular mass of the outflow in the overdense region, NW and far W clouds, and northern stripe (all included) is 1.8$\times$10$^{6}$--5.1$\times$10$^{6}$ M$_{\odot}$, resulting in a molecular outflow rate of 3.6--12.3 M$_{\odot}$ yr$^{-1}$.  The strictest lower limit for the molecular mass and molecular mass outflow rate, which only includes the northwestern cloud (an unambiguous outflow feature) are 1.5$\times$10$^{5}$  M$_{\odot}$ and 0.35 M$_{\odot}$ yr$^{-1}$, respectively. The values within these ranges are generally comparable to our adopted star formation rate of 4.7 M$_{\odot}$ yr$^{-1}$.  However, any outflowing material is likely reaccreted due to its low velocity compared to the escape velocity of Circinus.  
\par
6)  The molecular outflow rate is lower than for AGN-driven molecular outflows presented in the literature (e.g.\ \citealt{2014A&A...562A..21C}).  However, the existing literature sample is biased towards strong and massive outflows, with the result for Circinus perhaps being more typical for most galaxies.   
\par
7)  The kinetic power and momentum rate of the outflow are lower than those presented in \citet{2014A&A...562A..21C} for AGN-driven molecular outflows (especially given its high L$_{AGN}$ to L$_{bol}$ ratio), although it fits the predictions of some linear relations presented in that work (e.g.\ P$_{kin,OF}$ vs.\ L$_{AGN}$). When considering $v\dot{M}_{H_{2}, OF}$ compared to $L_{bol}$, Circinus is more consistent with the starburst galaxies within that sample.

\section{Acknowledgments}\label{acknowledgments}

First, we thank the referee for constructive comments and feedback that led to improvement of this manuscript, especially in its clarity.  This paper makes use of the following ALMA data: ADS/JAO.ALMA\#2013.1.00247.S . ALMA is a partnership of ESO (representing its member states), NSF (USA) and NINS (Japan), together with NRC (Canada), NSC and ASIAA (Taiwan), and KASI (Republic of Korea), in cooperation with the Republic of Chile. The Joint ALMA Observatory is operated by ESO, AUI/NRAO and NAOJ.  This research made use of APLpy, an open-source plotting package for Python hosted at http://aplpy.github.com.  Based on observations made with the NASA/ESA Hubble Space Telescope, obtained [from the Data Archive] at the Space Telescope Science Institute, which is operated by the Association of Universities for Research in Astronomy, Inc., under NASA contract NAS 5-26555. These observations are associated with program \#U4IM0101R.  Based on data products from observations made with ESO Telescopes at the La Silla Paranal Observatory under programme ID 094.B-0321.  JMDK gratefully acknowledges financial support in the form of a Gliese Fellowship and an Emmy Noether Research Group from the Deutsche Forschungsgemeinschaft (DFG), grant number KR4801/1-1.  EPF acknowledges funding through the ERC grant ``Cosmic Dawn". We thank I-Ting Ho, Filippo Fraternali, and Kazimierez Sliwa for for stimulating discussions concerning outflows and molecular chemistry. We thank C. Cicone for permission to reproduce figures originally published in \citet{2014A&A...562A..21C}. We thank A. Marconi for discussion concerning the status of work related to the MUSE data. We also thank R.G. Sharp for supplying optical IFU data of Circinus originally published in \citet{2010ApJ...711..818S}, which were helpful in our initial analysis but are not included here.

\bibliographystyle{apj}
\bibliography{circinus}

\nocite{2007A&A...474..837T}

\begin{appendix}
\section{Kinematic modeling of the Molecular Disk}\label{modeling}

\par
    In an attempt to extract emission with non-circular motions, we create a model galaxy cube.  We then subtract this model galaxy cube from the data cube in order to obtain residuals.  This approach, however, is very limited in the case of Circinus.  It is clear from recently observed IFU data of the ionized component (Figures~\ref{halpha_muse} and~\ref{channel_maps_muse}) that the outflow velocities substantially overlap with those of the main disk.  Further complicating this approach are the bar and spiral arms in Circinus, the former being aligned in the direction of the outflow, and the latter intersecting it at least twice (both as seen in projection).  With these issues in mind, a basic model galaxy is created in order to extract as much of the outflow as possible.

\par  
     We employ tilted-ring modeling \citep{1974ApJ...193..309R} to characterize the properties of the main disk using the Tilted Ring Fitting code (\textrm{TiRiFiC}, \citealt{2007A&A...468..731J}).  The model galaxy is comprised of a series of concentric rings for each of which we specify the surface brightness, inclination, position angle, rotational velocity, velocity dispersion, scale height, systemic velocity, and central position.  Thus, features such as warps, flares, bars, and spiral arms can be modeled.  Due to asymmetries, we model the approaching and receding halves of the galaxy separately.  A more detailed description of the modeling process can be found in \citet{2012ApJ...760...37Z}, \citet{2013A&A...554A.125G}, and \citet{2015ApJ...808..153Z} among others.

\par
   Initial estimates for the surface brightness are based on the distribution along the major axis.  The rotation curve is determined using the terminal side of the major-axis position-velocity (lv) diagram (Figure~\ref{lvplot}).  These parameters are continuously refined throughout the modeling process.  The velocity dispersion is gauged by the spacing of the contours on the terminal sides of all position-velocity diagrams and is found to be 15 km s$^{-1}$ for the main disk, but closer to 45 km s$^{-1}$ in the central few arcseconds.  (Recall that the higher dispersion at the center is likely due to the presence of a bar, thus it is not included in the models themselves.)  Several values for the exponential scale height are tested throughout modeling, including scenarios in which the scale height varies radially (although such variations are deemed unnecessary for a good fit).  A final value of 4" (80 pc in projection) is used. The inclination is initially assumed to be 65$^{\circ}$.  Additional values ranging from 50$^{\circ}$ to 70$^{\circ}$ are tested, but 65$^{\circ}$ is found to be the best fit throughout the disk. Position angles between 200$^{\circ}$ and 230$^{\circ}$ are tested, with an optimal value of 210$^{\circ}$.  Although, in the regions observed, the position angle is not best fit with a single value, but with fluctations due to the bar and spiral arms (or a bar and spiral arms superimposed as we have done in our final model).  

\par
\begin{figure}
\includegraphics[width=80mm]{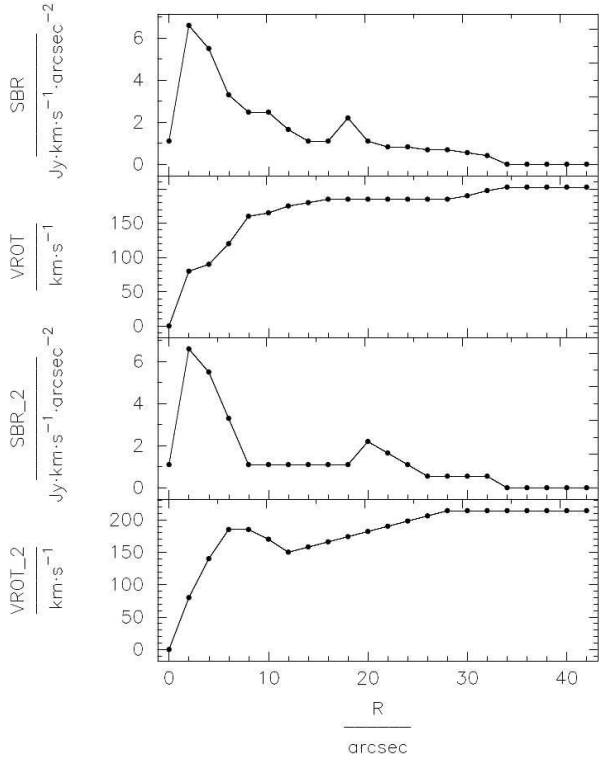}
\caption{\textit{Rotation curves and surface brightness for the models prior to adding the bar and spiral arms as they cannot be quantified over all radii in the same manner. The approaching half is shown in the top panels, and the receding half is shown in the bottom.} \label{1c_2halves}}  
\end{figure}
\begin{figure*}
\begin{centering}
\includegraphics[width=150mm]{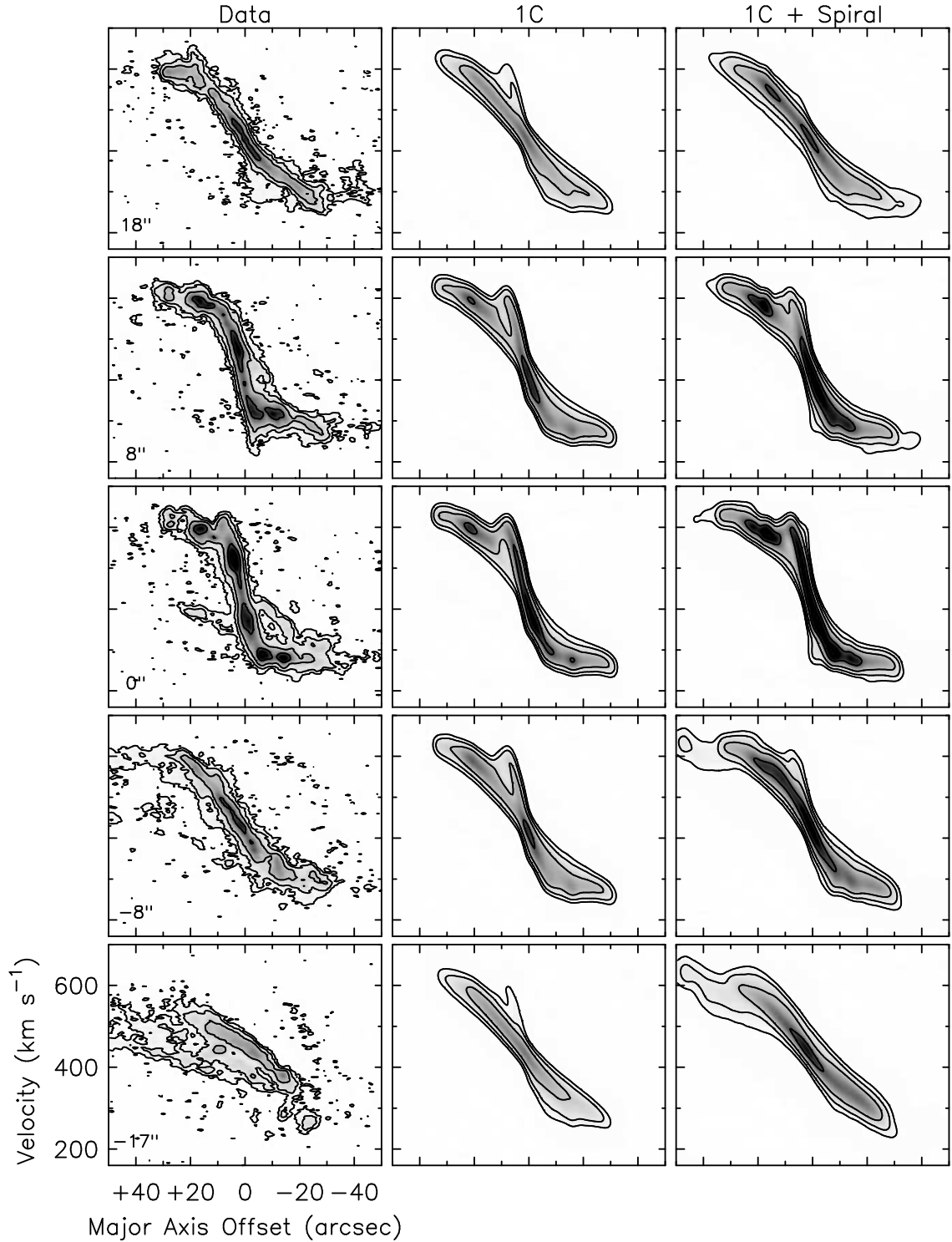}
\caption{\textit{Position-velocity diagrams parallell to the major axis (lv diagrams) of the data, 1C, and 1C + spiral models.  The material shifted towards systemic is almost entirely due to the streaming motions in the spiral arms.  The blue-shifted feature at 20" along the major axis corresponds to the southern disturbance.  The component shifted closer to systemic from 0" to --20" along the major axis corresponds to the northern stripe.  } \label{lvplot}}  
\end{centering}
\end{figure*}
\begin{figure*}
\begin{centering}
\includegraphics[width=80mm]{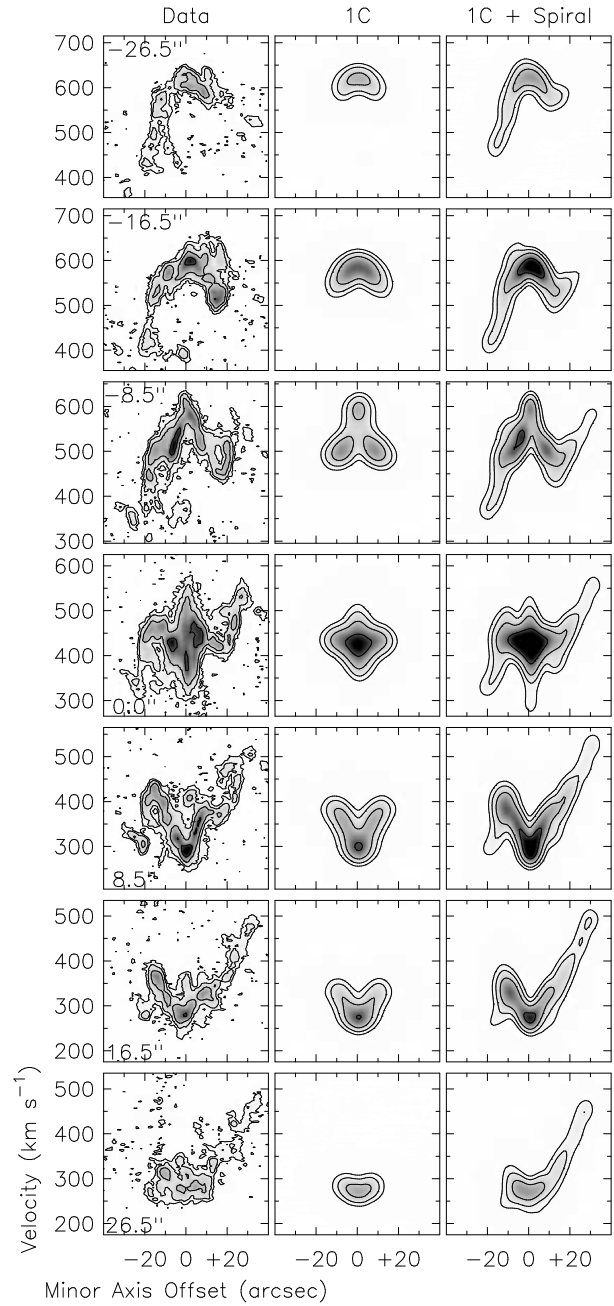}
\caption{\textit{Position-velocity diagrams parallel to the minor axis (bv diagrams) of the data 1C, and 1C + spiral models.  The stretching of features in velocity outside of $\pm$20" is due to large streaming motions in the spiral arms.  Because of the non-circular velocities within the bar and (primarily outer) spiral arms, unambiguous outflow signatures cannot be extracted directly from these models.} \label{bv}}  
\end{centering}
\end{figure*}

    From the start it is clear that the fitting process is complicated by the presence of a bar and spiral arms extending from the center to the edge of the disk.  Disk parameters for the final models are shown in Figure~\ref{1c_2halves}.   The first (1C) model consists of only a single disk of constant inclination and position angle (column two of Figures~\ref{lvplot} and~\ref{bv}).  The second model (1C + spiral) consists of the same 1C disk model, but with gaussian distortions to reproduce the bar and spiral arms (column three of Figures~\ref{lvplot} and~\ref{bv}).  We retain symmetry whenever possible (i.e. modeling one side, and then flipping it 180$^{\circ}$ to fit the other half before making as few alterations as possible). As may be expected, the bar and spiral arms appear to have some non-circular velocities.  The orientation of the bar (i.e.\ directly along the direction of the expected outflow) makes it impossible to unambigously distinguish between what is part of the bar and what is outflowing material.  The innermost spiral arms exhibit primarily disk-rotation.  However, the velocities of the outermost spiral arms extending beyond the main disk are observed to be close to systemic for more than 20" (400 pc) along the length of each.  If the rotation curve is matched to the main disk, a substantial radial inflow of 100--150 km s$^{-1}$ within the outer arms is the only way to reproduce the velocity field.  Unfortunately, the precise details of these radial motions, rotational velocities, pitch-angle, inclination are degenerate to some degree.  The northwestern arm intersects the location of the ionized outflow.  With degenerate parameters for the spiral arms, it is not possible to reliably extract the molecular outflow in this region. 

\par
We obtain initial residuals and find indications of more material present in the overdense region, but nothing conclusive.  We then use a more agressive approach and oversubtract the models in order to eliminate virtually all material with disk-rotation (thus also removing any outflow material that shares disk velocities).  This approach does not yield any unambiguous detection of outflowing molecular gas either (Figure~\ref{residual_moments_all}).  While a similar approach has been applied to other galaxies (e.g.\ \citealt{2014A&A...567A.125G}, \citealt{2016arXiv160703674P}), yielding clear outflow signatures, it is not well-suited in the case of Circinus because of the shared disk and outflow velocities, as well as the orientation and non-circular motions of the bar and spiral arms.   

\par
The quality of the models is sufficient to demonstrate the extent to which this technique may be used to extract the outflow in Circinus.  There are some minor imperfections in the spiral arms, but their fits would be degenerate in any case.  It may be fruitful to perform an exhaustive modelling analysis in order to further constrain the streaming motions in the spiral arms and kinematics of the disk itself, but such an analysis is unrelated to the outflow and beyond the scope of this paper.  

\end{appendix}

\end{document}